\documentclass[sigconf]{acmart}

\AtBeginDocument{%
  \providecommand\BibTeX{{%
    \normalfont B\kern-0.5em{\scshape i\kern-0.25em b}\kern-0.8em\TeX}}}

\copyrightyear{2023}
\acmYear{2023}
\setcopyright{rightsretained}
\acmConference[CHI '23]{Proceedings of the 2023 CHI Conference on Human Factors in Computing Systems}{April 23--28, 2023}{Hamburg, Germany}
\acmBooktitle{Proceedings of the 2023 CHI Conference on Human Factors in Computing Systems (CHI '23), April 23--28, 2023, Hamburg, Germany}\acmDOI{10.1145/3544548.3581273}
\acmISBN{978-1-4503-9421-5/23/04}

\usepackage{graphicx}  
\usepackage{url}
\usepackage{booktabs}  
\usepackage{caption}
\usepackage{subcaption}
\usepackage{booktabs}
\usepackage{tabularx}

\graphicspath{{figures/}{pictures/}{images/}{./}}

\def\add#1{#1}
\def\remove#1{\unskip}

\begin{document}

\title[Collaborative Practices and Tools of UX Practitioners]{Understanding Collaborative Practices and Tools of Professional UX Practitioners in Software Organizations}


\author{K. J. Kevin Feng}
\affiliation{%
  \institution{University of Washington}
  \city{Seattle}
  \country{USA}}
\email{kjfeng@uw.edu}

\author{Tony W. Li}
\affiliation{%
  \institution{University of Washington}
  \city{Seattle}
  \country{USA}}
\email{tonywli@uw.edu}

\author{Amy X. Zhang}
\affiliation{%
  \institution{University of Washington}
  \city{Seattle}
  \country{USA}}
\email{axz@cs.uw.edu}

\renewcommand{\shortauthors}{Feng et al.}

\begin{abstract}
  User experience (UX) has undergone a revolution in collaborative practices, due to tools that enable quick feedback and continuous collaboration with a varied team across a design’s lifecycle. However, it is unclear how this shift in collaboration has been received in professional UX practice, and whether new pain points have arisen. To this end, we conducted a survey ($N=114$) with UX practitioners at software organizations based in the U.S. to better understand their collaborative practices and tools used throughout the design process. We found that while an increase in collaborative activity enhanced many aspects of UX work, some long-standing challenges---such as handing off designs to developers---still persist. Moreover, we observed new challenges emerging from activities enabled by collaborative tools such as design system management. Based on our findings, we discuss how UX practices can improve collaboration moving forward and provide concrete design implications for collaborative UX tools.
\end{abstract}




\maketitle

\section{Introduction}

Today, the field of user experience (UX) is viewed as a cornerstone in the human-centered design of interactive systems \cite{gray2015, norman2016, nielson2017}. UX typically refers to principles and practices that enhance an end-user's experience with products, services, and companies \cite{nngroup-ux}, and it has seen rapid growth in both the number of practitioners in the field \cite{nielson2017} as well as adoption within a wide range of organizations \cite{gray2015}. Interest in UX jobs saw a 289\% increase in 2020 alone \cite{state-ux-2022}, while more companies are reaching a sufficiently high level of UX maturity to hire managers to lead entire UX teams \cite{nielson2017}. 
Furthermore, UX practitioners (UXPs) are increasingly working with technical, non-UX stakeholders such as developers and data scientists, as the incorporation of complex technologies such as machine learning necessitates closer and lengthier UX collaboration \cite{yang2018}.

Alongside this growth in size and stature, the field has seen a major shift towards collaborative UX tools \cite{2021-tools}. Popular tools now boast robust features for synchronous and asynchronous collaboration as a core product offering \cite{figma, xd, sketch-collab, framer, miro} to enable teams of UX practitioners (UXPs) to work together on UX challenges of higher complexity and scale. It was not always this way. Sketch, a Mac-only interface design tool that dominated the field throughout much of the 2010s \cite{2017-tools}, was once a native app that forced UXPs to work locally and share files through email. In 2016, Figma released its browser-based interface design tool enabling real-time collaboration on a shared canvas \cite{figma}, and other tools (including Sketch) soon followed \cite{xd, sketch}. In addition to eliminating the versioning and synchronization issues that inevitably arose from working locally and manually sharing design files, these tools revolutionized collaboration among UXP teams by opening up new potential for collective ideation and critique, as well as tighter feedback loops. More broadly, they democratized UX for non-UXPs and even nurtured a community of UXPs to share reusable templates and tool extensions \cite{figma-community}.


These shifts lead us to consider whether there may be new challenges and needs for UXPs engaged in collaboration in light of the major changes in collaboration practices of UXPs and changes in collaborative tools. \add{Prior work suggests that increased collaboration with technical stakeholders may exacerbate UX challenges such as conceptual gaps and coordinative breakdowns during periods of handoff \cite{maudet2017breakdowns, yang2020difficult}. While this line of work surfaces important pitfalls, those pitfalls are not discussed alongside the recent additions of collaborative capabilities in common tools, which have since become an integral part of UX practice \cite{2021-tools}. A limited set of previous work has catalogued the changing landscape of UX tools in recent years \cite{2017-tools, 2021-tools} and how collaborative features impact tool adoption \cite{stolterman2012tools} or support specific tasks such as ideation \cite{inie2020}. These works, however, do not provide a high-level overview of the diverse challenges UXPs may face within such tools. Additionally, besides user complaints in tool forums \cite{xd-forum}, there has been little empirical work exploring new challenges induced by collaborative tools. Literature outside of UX provides strong indication that such challenges exist---prior work outlined forms of coordinative friction that arose with the introduction of collaborative document editors \cite{wang2017writing} and code editors \cite{goldman2011}. Yet, it would be unwarranted to generalize these findings to UX given the prevalence of domain-specific representations and processes \cite{maudet2017breakdowns, yang2018}.}

Despite the rise in both UX teams and UXP and non-UXP collaboration, there is surprisingly little empirical understanding of how these collaborations play out among different roles across stages of the design process. \add{This understanding is vital in a post-COVID-19 era. As organizations and employees rapidly adopt hybrid work models \cite{mckinsey}, disparate roles are likely to be physically distributed and their work will increasingly be mediated by collaborative tools. Indeed, usage of Figma and Miro increased sharply throughout the COVID-19 pandemic and remain at unprecedented levels \cite{2021-tools, 2019-tools}. New workflows may have also emerged from usage of collaborative tools that pre-pandemic literature may fail to capture. Thus, the time is ripe to obtain an updated understanding of how UXPs collaborate, the tools they use, and in particular how these two intersect and influence each other. We pose 4 main research questions with these goals in mind:}

\begin{itemize}
    \item \textbf{RQ1:} With whom do UXPs collaborate at each stage of the design process?
    \item \textbf{RQ2:} What tools and tool-based collaboration strategies are UXPs using throughout their workflows?
    \item \textbf{RQ3:} What practices and challenges arise when UXPs \remove{work with technical collaborators such as} \add{hand off their work to} software developers and \remove{data scientists} \add{ML practitioners}?
    \item \textbf{RQ4:} What are UXPs' current expectations and practices around reusing design components and managing design systems in collaborative UX tools?
\end{itemize}

In this paper, we answer these questions by conducting a descriptive survey with 114 UXPs employed at software organizations---organizations that actively maintain and/or build new software---based in the U.S. to understand their collaborative practices, tools, and relationships between the two. \add{We use a survey as our research instrument of choice due to its ability to simultaneously capture high-level trends and probe into specific practices, as demonstrated by prior survey-based work \cite{2019-tools, 2021-tools, zhang2020data, serban2020adoption}.} We use the term ``tool'' to specifically refer to software applications (e.g., Adobe Illustrator), software artifacts (e.g., code, spreadsheets), and physical objects (e.g., pen and paper) that aid and enhance UXPs' work. 
We qualitatively analyzed free responses from our survey to extract themes and insights, tallied categorical responses, and applied quantitative methods to identify significant differences where relevant. A summary of our most salient findings is as follows:

\begin{enumerate}
    \item UXPs collaborate with a wide range of roles, but most frequently with other UXPs and product managers. However, UXPs will typically be the ones to initiate collaborations with non-UXPs.
    
    \item Most UXPs primarily use a single, all-in-one collaborative UX tool such as Figma to perform functions and tasks across all stages of the design process.
    
    \item UXPs use text-based strategies, such as comments and on-design annotations, while sharing their work with collaborators but often prefer verbal methods such as meetings and voiceovers. UXPs also employ more visuals and storytelling techniques when presenting their work to non-UXPs compared to fellow UXPs.
    \item \remove{Working with technical stakeholders remains}\add{Handoffs remain} a tricky challenge for UXPs due not only to persistent problems around knowledge gaps and design-development divergence, but also problems unique to collaborative UX tools such as overcommunication of irrelevant design details to developers.
    \item While greater collaboration on design systems within collaborative UX tools has popularized sharing and reusing design components, new issues around design system management have arisen that may exacerbate coordinative breakdowns: ambiguity of ownership, lack of communication about changes, overly frequent changes, and more.
\end{enumerate}



Through the lens of our findings, we reflect on the state of collaborative practices of UXPs in light of major shifts in tools in the last several years. We argue that today's UXPs use tools that act as a connective ``hub'' of  activity across an entire team, where cross-functional stakeholders can collectively manage information from a single workspace. Yet, many challenges when working with technical stakeholders still persist, and new sources of friction emerge from activities made possible by this collaborative shift, such as management of scalable design systems. We conclude by recommending advancements in collaborative techniques and tools that do not merely \textit{offer} opportunities for collaboration but also \textit{enrich} collaborative experiences within and around UX. 



\section{Background and Related Work}
Collaboration in UX draws upon a multitude of traditions in computer-supported collaborative work and human-centered design. Here, we review relevant past works in collaboration among (non-UX) designers, UX practices, and collaborative tools to better set the stage for our work. 

\subsection{Collaboration Among Studio Designers}
Researchers have been interested in how designers collaborate well before UX was popularized as a field. Works that did not consider software-based UX focused primarily on architectural and industrial design studio environments \cite{bellotti1996, chiu2002, elsbach2013, vyas2013creative, schmidt2002coordinative}, as the studio was considered central to design work and education \cite{ochsner2000behind, vyas2013creative}. As computers made their way into design studios, tradeoffs surfaced between physical and remote software-based collaboration \cite{bellotti1996}. Indeed, the lack of interactivity in early computing tools presented obstacles to digitally communicating sophisticated designs and prompted further research on how software tools can help with these obstacles \cite{kolarevic2000, budd1999, lee2009, chiu2002}, including virtual asynchronous design studios ~\cite{kolarevic2000, budd1999}. Years later, when virtual work practices may be forced upon many in times such as during the COVID-19 pandemic, Lee et al.~\cite{lee2009} showed that designers considered a distributed collaboration workflow to be more engaging than a face-to-face one because they were forced to put more thought into communicating effectively. Though the digitization of studio workspaces is not met without resistance, it unlocks flexible, accessible, and efficient collaboration workflows for designers \cite{iranmanesh2021mandatory, jones2021longitudinal}.

While many studies of studio design collaboration were conducted just as computers were becoming mainstream office tools, their insights 
can be seen integrated into modern productivity tools. Multimedia access for instructional content explored by ID-Online \cite{budd1999} is widely available in Canvas \cite{canvas}, a popular class management system used by universities worldwide. Separation of workspaces and discussion areas is available in general-purpose project management software such as Jira \cite{jira} and Asana \cite{asana}. Shared databases with asynchronous contribution tracking, such as the one implemented by Kolarevic et al. \cite{kolarevic2000}, are now ubiquitous in cloud-powered productivity hubs such as OneDrive \cite{onedrive} and Google Workspace \cite{google-drive}. 

Although modern collaboration software appears to address some past desiderata, it is unclear whether they result in new collaborative challenges of their own. Additionally, collaborative practices have been studied extensively in traditional studio settings, but those same practices may not be preserved in contemporary UX workflows for software interfaces. We tackle these questions in our study.

\subsection{\remove{Multiplayer Collaboration}\add{Multi-User Collaborative Authoring} Tools} %
\label{s:rel-works-tools}
Tools that enable \remove{collaboration} \add{collaborative authoring}, particularly ones that allow simultaneous multi-user interactions, have captured the interests of researchers in diverse domains, including software engineering \cite{goldman2011, hattori2010syde, warner2017scaffolding, liveshare, bani2008integrating, ghorashi2016jimbo}, data science \cite{zhang2020data, wang2019, wang2020, patterson2017dataflow, smith2017featurehub, mendez2019toward}, document editing \cite{lee2022coauthor, ginige2010collaborative, jung2017docs, teevan2016writing, kim2021winder}, entertainment \cite{lu2021, shakeri2021saga}, and geospatial analysis \cite{fechner2015}, many of whom observed novel or unforeseen issues arise from the collaborative features. \add{For example, Goldman et al. built a collaborative, web-based Java IDE called Collabode \cite{goldman2011}, but found upon evaluation that programmers' code produced runtime errors for collaborators despite it being free of compilation errors. Wang et al.~\cite{wang2017writing} conducted an interview study ($N=30$) to investigate rationales for writing practices using collaborative writing tools such as Google Docs. They found that despite the rich suite of co-editing features, users were reluctant to fully embrace them due to concerns of accountability, credit of contribution, and loss of privacy.} In general, dynamics of performing tasks in collaborative tools may be different than performing those same tasks in an isolated environment \cite{erickson2000social} and deserve scrutiny during evaluation.

\add{While prior work in this area yields rich findings, there are still numerous unanswered questions when it comes to collaboration in the UX domain. Findings in non-UX domains are not guaranteed to generalize to UX work given representational and procedural differences \cite{maudet2017breakdowns, yang2018}, while work studying collaborative UX tools are remarkably limited, possibly due to the tools' recent adoption. Indeed, when Jung et al. \cite{jung2017docs} studied small-team remote collaboration in design challenges, they used Google Docs as their tool of choice due to its overwhelming popularity at the time rather than a specialized design tool such as Figma. Furthermore, prior evaluations of collaborative tools primarily probed practitioners' experiences working with others in their own domain \cite{goldman2011, wang2019, wang2020, kim2021winder, jung2017docs}. However, studies of UX in professional settings revealed that UXPs frequently interface with colleagues outside of UX \cite{gray2015, gray2016}, as small or even single-person UX teams were common. We account for this by investigating both inter- and intra-domain collaboration in our survey.}

\remove{UX tools have also recognized the need for collaboration support.} \add{UX tools in industry have charged forward with releasing collaborative features, despite little academic work in this area.} Figma \cite{figma} is a browser-based UI design and prototyping tool that allows multiple users to work in real-time and has gained significant popularity since its launch in 2016 \cite{2021-tools}. Its popularity led to native tools such as MacOS's Sketch \cite{sketch} and Adobe XD \cite{xd} also enabling real-time collaboration. Social affordances for co-presence, such as multiple cursors on a single canvas, are also popular in other whiteboarding and ideation tools, such as Miro \cite{miro} and Mural \cite{mural}, as well as platforms to share community assets, such as Figma's Community \cite{figma-community}. \add{Interestingly, community-building in design tools has been shown by previous work to be a catalyst for tool adoption among practitioners \cite{stolterman2012tools}. These collaborative tools have enabled now-common UX practices around design system management \cite{churchill2019scaling, moore2020systems}, where modular design components can be reused across an entire organization to ensure consistency, and developer handoffs \cite{figma-handoff}, where the designer passes their work to developers for code implementation. Consequently, design systems have been further formalized. Moore et al.~\cite{moore2020systems} identified three core components of design systems---a \textit{design philosophy} for establishing an overall vision for the system, \textit{interaction patterns} that compose the system's core user experience, and \textit{content format} that helps generate in-system content (e.g., images, data formats) for particular use cases. Despite this formalization, there has been no work, to our knowledge, investigating UXPs' experiences of collaboratively managing design systems.}

Given previous literature and industry trends, attention to tool-based collaboration support is clearly high. \add{The landscape of UX tools has shifted such that collaboration is now placed at the forefront of their experiences, but we do not have an updated understanding of how this shift influences the way designers work, or what new collaborative friction, if any, has arisen from new collaborative features.} \remove{Despite the increased number of collaboration features in tools used in UX, there is little exploration into how these features influence the way designers work, or what new collaborative friction, if any, has arisen from these features. For example, UX practitioners managing design systems in a tool such as Figma may be required to consider component reusability issues, something that may not have been top-of-mind in localized workflows.} \remove{The novelty of collaborative capabilities in UX tools, as well as u} \add{U}nforeseen circumstances that demand high usage of them---such as a global pandemic---may \remove{be contributors} \add{also contribute} to the current gap in literature that directly tackle these questions. We bridge this gap \remove{with our work} \add{by capturing both a high-level understanding of the current status quo in UX, as well as specific UX practices where increased collaboration may be contentious}. \add{While we could have conducted an interview study, scaling up the number of participants was important for us to obtain a higher-level view of the field, so we opted for a survey. However, we still wanted participants to share specific, nuanced experiences with us, so we designated many survey questions as open-ended responses. The resulting \textit{descriptive survey} \cite{macfarlane1996conducting} thus became our study instrument of choice.}

\subsection{\remove{UX Work Practices}\add{Characterizing UX Work}} 

\add{The growth of UX as a field \cite{nielson2017} has prompted researchers and practitioners alike to more formally characterize UX work. Below, we review some relevant works that helped inform the \textit{lingua franca} of modern UX, which we leverage in our survey.}

\remove{Don Norman is widely credited with coining the term ``user experience'' in 1993 \cite{nielson2017} in an attempt to formalize a system that covers all aspects of a person's experience with a personal computer, including the interface, graphics, physical interactions, and the paper manual \cite{norman2016}.   
Since then, the number of employees trained in UX, the number of companies practicing UX, and the number of countries investing in UX initiatives have all grown steadily, and the term has become ubiquitous in the technology industry and beyond \cite{nielson2017}. Despite (and perhaps partly due to) UX's proliferation, there is still no firm consensus on one definition of UX. Efforts in previous literature include a ``UX manifesto'' \cite{law2008}, a white paper \cite{roto2011}, surveys \cite{law2009, lallemand2015}, and literature reviews \cite{gomez2019, berni2021}. Even if a definition is reached, it may be fleeting due to UX's constant evolution and multidisciplinary nature \cite{lopez2019ux}. For the purposes of this paper, we use UX to broadly refer to the experience of interacting with software UIs, excluding the hardware and print experiences outlined by Norman in his original definition.}
\remove{UX processes are arguably more concrete than the definition of the field.} The British Design Council popularized the double diamond design process in 2004 and describes it using 4 distinct phases: Discover, Define, Develop, and Deliver \cite{design-council-2005}. \remove{The process draws upon an older discipline of user-centered design (UCD) \cite{courage2005} while incorporating other professional practices for designing effective digital products and services, including information architecture, content strategy, and user interface (UI) design \cite{getto2016}.} The double diamond has been adopted as a standard UX workflow by practitioners in industry and beyond \cite{yang2018, clune2014, reunanen2020, hall2013}. Activities commonly associated with each phase of the process are listed in Table \ref{t:activities}. 

\begin{table*}
    \centering
    \begin{tabular}{m{2cm} m{6cm} m{6cm}}
     \toprule
     \bf Phase & \bf Description & \bf Common Activities\\
     \midrule
     Discover & Comprehending and questioning the design problem and understanding needs of potential users. & Conducting interviews, questionnaires, field observations\\ \midrule
     Define & Converging possibilities through interpreting and aligning research findings to project objectives. & Synthesizing data from interviews, finalizing the design problem, defining personas \\ \midrule
     Develop & Exploring multiple directions with the goal of revealing attributes of an effective solution. & Sketching wireframes, conducting usability tests, iterating on low-fidelity prototypes \\ \midrule
     Deliver & Presenting a single solution and preparing it for launch and/or further iteration & Creating high-fidelity prototypes, writing pitch documents, publishing user stories \\ 
    \bottomrule
    \end{tabular}
    \caption{Outline of phases in the double diamond design process.}
    \label{t:activities}
\end{table*}

\begin{figure*}
    \centering
    \includegraphics[width=0.7\textwidth]{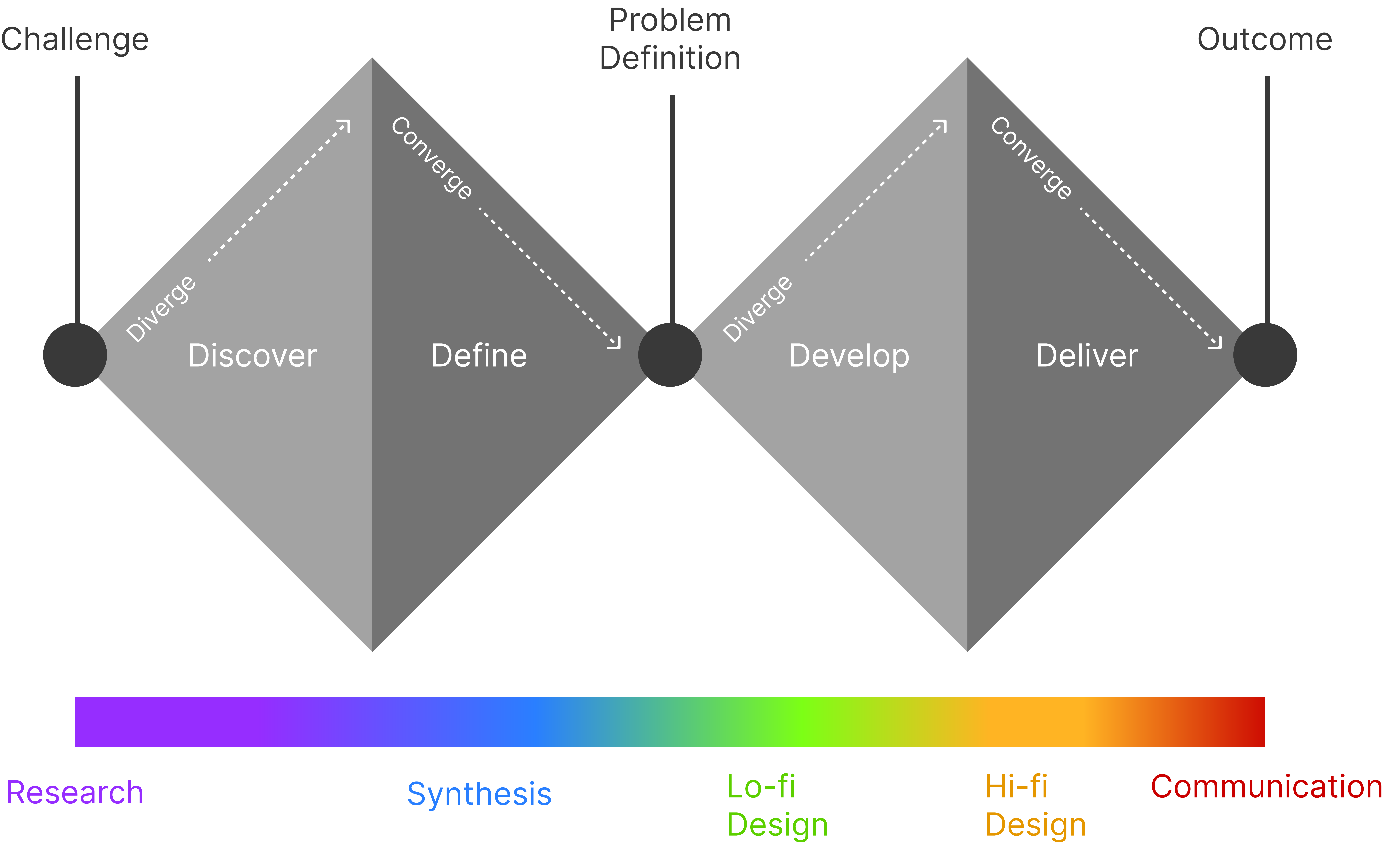}
    \caption{Common categories of UX activities mapped onto the double diamond design process \cite{design-council-2005}.}
    \label{f:dd}
\end{figure*}

Although not all UX workflows may follow the double diamond, it is a widely adopted framework with distinct activities and methods across its stages; hence, we take inspiration from it to scaffold our analysis. To make the stages more concrete for participants, we mapped its stages to 5 common categories of UX activities \cite{ux-stages} that loosely correspond to these stages---research, synthesis, low-fidelity design, high-fidelity design, and communication. Our mapping is described in Fig. \ref{f:dd}, where we use a gradient to show loose correspondence.

\add{The increasing rate of UX adoption within companies \add{\cite{nielson2017}} has led to studies of UX in organizational contexts. Gray et al. ~\cite{gray2015} noted that despite companies' enthusiasm to embrace UX, company culture can still be hostile to the consideration of new UX perspectives, which leads to frustration from UX designers. A select number of works examine topics in UX collaboration with particular interest in integration with technical stakeholders through designer-developer handoffs \cite{leiva2019enact, maudet2017breakdowns, walny2020handoff}, Agile software development processes \cite{jurca2014, kuusinen2012, ovad2015, liikkanen2014}, and collaboration with data scientists and AI engineers \cite{yang2018, subramonyam2021towards}, in addition to other UX practitioners for research synthesis \cite{kuang2022data}. Frishberg and Convertino argue for the importance of UX collaboration with non-UX parts of the organization, and recognize collaborative differences in UX teams situated in corporate departments versus design agencies \cite{frishberg2020}. We employ this observation in our work by capturing information about how participants' UX teams are situated within their larger organization, while excluding design agencies from our scope.}
 
\add{UX's proliferation as a professional practice also led to an expansion of UX-related roles. As researchers, roles---often communicated through job titles---provide us with crucial information on the nature of UX work participants engage in, as well as whether they work in UX in the first place.} \add{Indeed,} contemporary UX job titles can reflect ownership of different areas of the double diamond process \cite{lauer2014postings, putnam2012professions}, \remove{which}\add{but} may vary based on organization \cite{gray2016}. That said, many UX titles are still used fluidly and are ambiguous in their roles---these include Interaction Designer, Experience Architect, UX/UI Front End Developer, among others \cite{lauer2014postings}. Even non-UX workers, such as software engineers, are beginning to take on select parts of the design process themselves if it is closely tied to their work \cite{gray2015}. We take this discourse on job titles into consideration in our survey by allowing participants to self-select their job function from 5 higher-level categories we observed from aforementioned work and recent UX job postings rather than aggregating by participants' job titles. Those 5 categories are research, design, engineering, writing, and other.

Despite current literature on UX\remove{design}, we still lack an updated understanding of UX\remove{designers} 
 \add{practitioners}' collaboration practices that holistically consider organizational factors, tools, and workflows. Changes in workplace culture due to the COVID-19 pandemic mean that some factors traditionally fostering collaboration, such as sitting next to each other \cite{jones2019}, may be more demanding to implement. Furthermore, it is not clear from previous work how UX designers collaboratively move between tasks in the double diamond design process, particularly in larger organizations where a designer may only work in a specific phase of the process. Our study serves to shed light on these ambiguities.


\section{Method}
The goal of our study is to explore collaboration patterns and tool use in UXPs working on software products. To accomplish this, we deployed a survey hosted on Qualtrics\footnote{\url{https://www.qualtrics.com/}} to UXPs working in U.S.-based organizations that actively created and maintained software. We focused on U.S.-based organizations to limit variations in regional UX practices and definitions that may permeate organizations with differing ``home bases'' \cite{lallemand2015}. While prior work investigated collaborative behaviours between industry UXPs in specific parts of the design process \cite{kuang2022data, inie2020, oleary2018charrette}, we observe collaboration patterns across the entire UX lifecycle, taking into consideration roles, team size, and common tools used. 

\subsection{Survey Questions}
\add{Our survey questions were informed by surveys from prior work that probed practitioners' technical practices within technology organizations \cite{zhang2020data, holstein2019fairness, wang2019}, as well as the authors' recent experiences working with UX practitioners in technology organizations. We also discussed our survey questions with two UXPs from our personal networks to refine or remove existing questions and add new ones.} The survey first collected some basic demographic information such as gender, years of experience working in UX and in their organization more broadly, UX team size and structure, and the stage(s) of the design process they were involved in. Participants identified a UX role they associated with most closely in their project from our list of 5 roles---designer, researcher, engineer, writer, and other. These UX roles were compiled by considering prior work in UX job titles \cite{lauer2014postings, putnam2012professions} and industry trends, such as the emergence of the UX writing role \cite{ux-writer}. We also defined a set of non-UX roles common in software organizations based on a similar list by Zhang et al. \cite{zhang2020data}. The non-UX roles are product manager, software developer, data scientist, domain expert, and sales/marketing/finance. Participants were then invited to recall a project they previously completed. With that recollection in mind, they answered multi-select questions on who they collaborated with (both UX and non-UX roles) throughout their UX workflows, along with free-response questions about tools, strategies, and challenges. 

Depending on whether participants created designs themselves (as opposed to just conducting research or writing code), they were asked a series of design-specific questions, from design system management, to developer handoff, to design environment customization. We focused on these areas as previous work indicated that these may be locations of \remove{pain points}\add{uncertainty and coordinative tension} in the UX design process \cite{churchill2019scaling, figma-handoff, pacheco2021improving}. We piloted the survey with members of our lab, which led us to modify the survey structure and clarify terminology in the questions based on feedback. Our full survey is available in our supplementary materials.

\subsection{Survey Distribution and Participants}
We sent our survey to our department and department alumni Slack channels related to HCI and design. We reached out to our personal connections in U.S.-based technology companies and asked them to pass along the survey to UXPs within their organization. We also posted the survey to design UX-related social media groups on Discord, LinkedIn, Facebook, Reddit, Blind, and Twitter (both on Twitter Communities as well as from authors' personal accounts). We began recruiting in late February 2022 and accepted our last response in early August 2022, when we reached data saturation. We leveraged snowball sampling by implementing a referral system where participants received an additional entry to our gift card raffle for every participant they referred who also completed the survey. 

We received valid survey responses from 114 participants. Responses were considered valid if the participant held a UX-related job title and self-identified as one of five UX roles we derived from previous academic and industry literature \cite{lauer2014postings, putnam2012professions, ux-titles-article}. All participants who contributed valid responses were entered into a raffle to win one of five \$50 USD Amazon gift cards. 

\begin{figure*}
    \centering
    \begin{subfigure}[h]{0.45\textwidth}
        \centering
        \includegraphics[width=\textwidth]{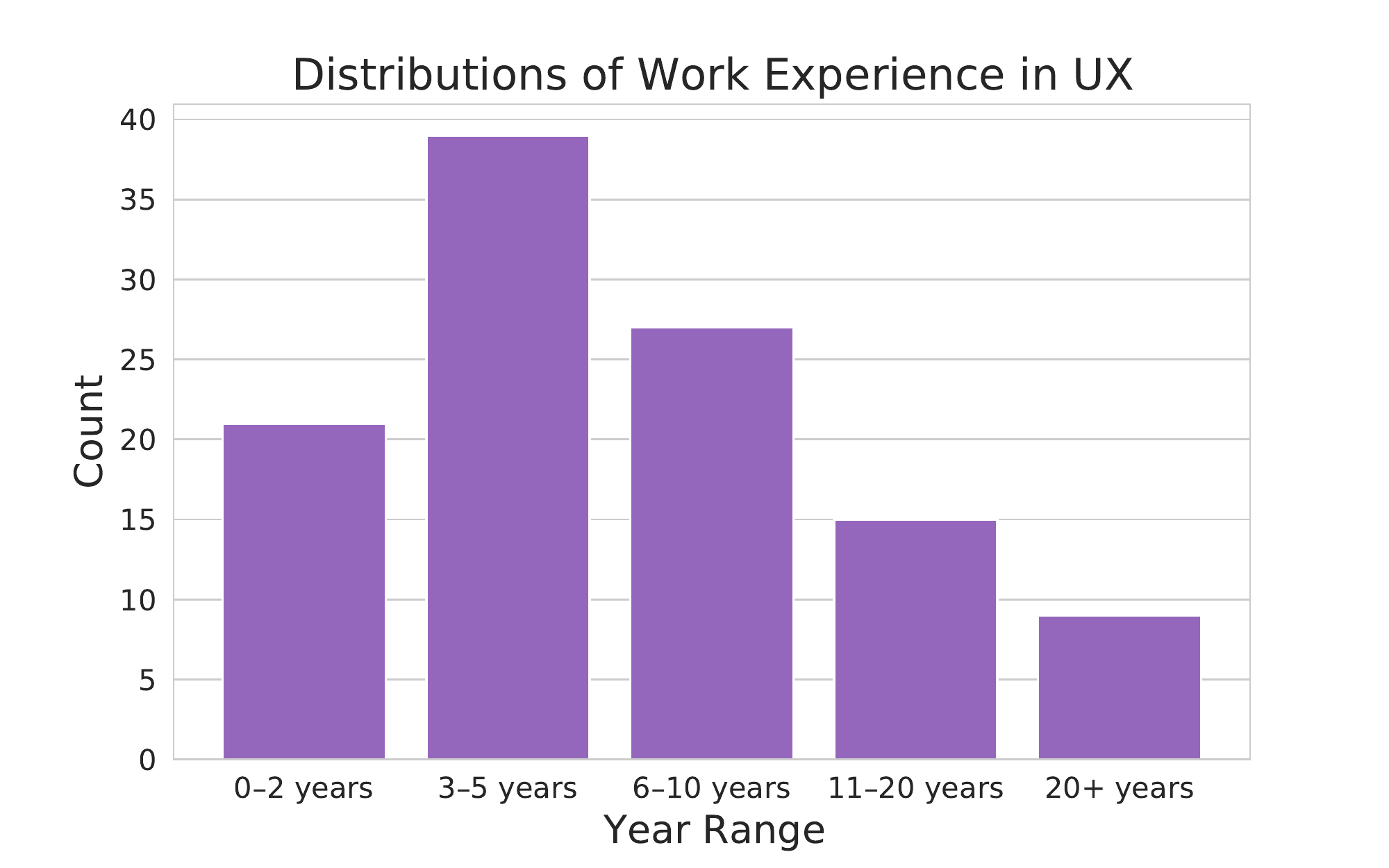}
        \caption{}
        \label{f:demo-1}
    \end{subfigure}
    \begin{subfigure}[h]{0.45\textwidth}
        \centering
        \includegraphics[width=\textwidth]{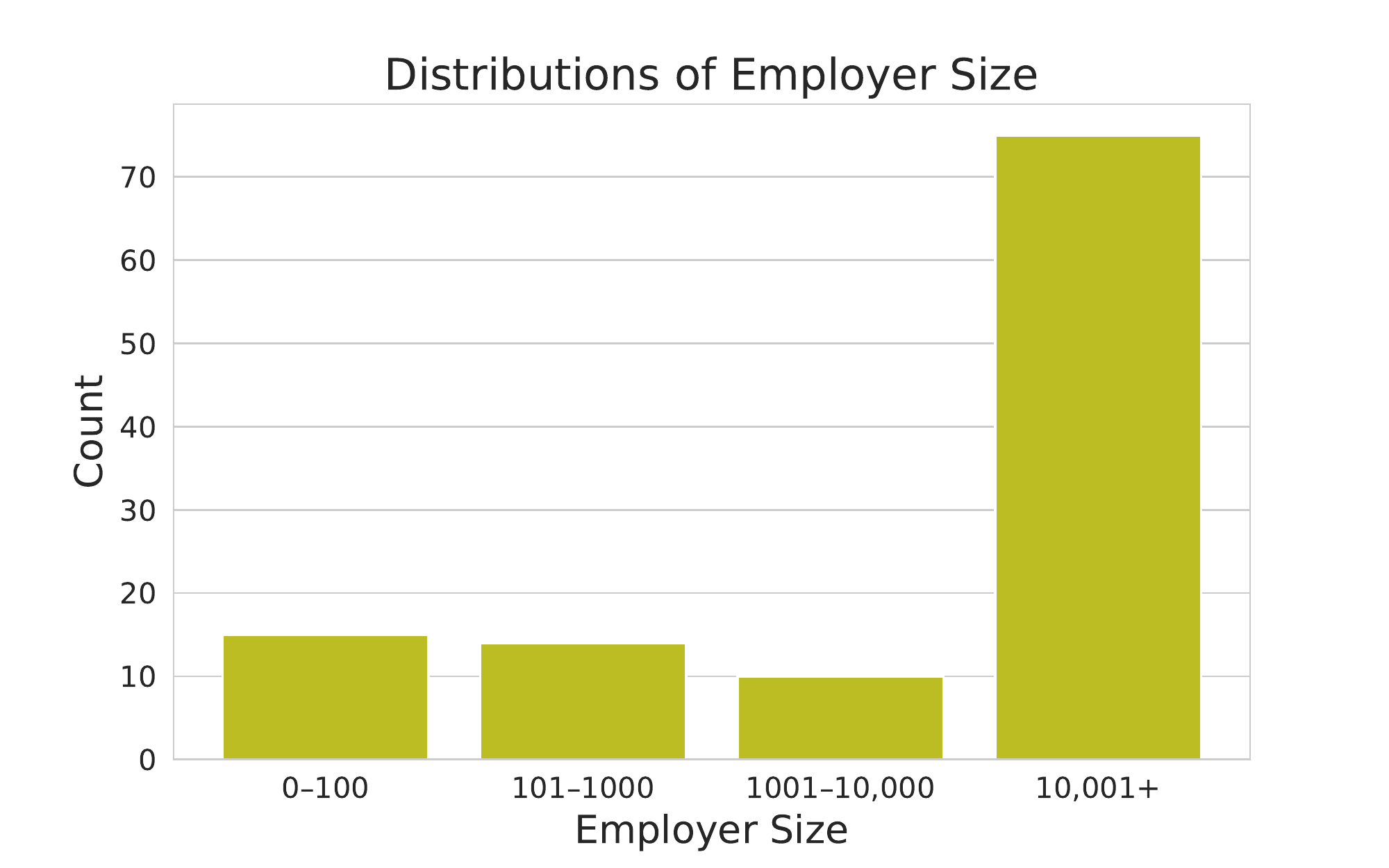}
        \caption{}
        \label{f:demo-2}
    \end{subfigure}
    \begin{subfigure}[h]{0.45\textwidth}
        \centering
        \includegraphics[width=\textwidth]{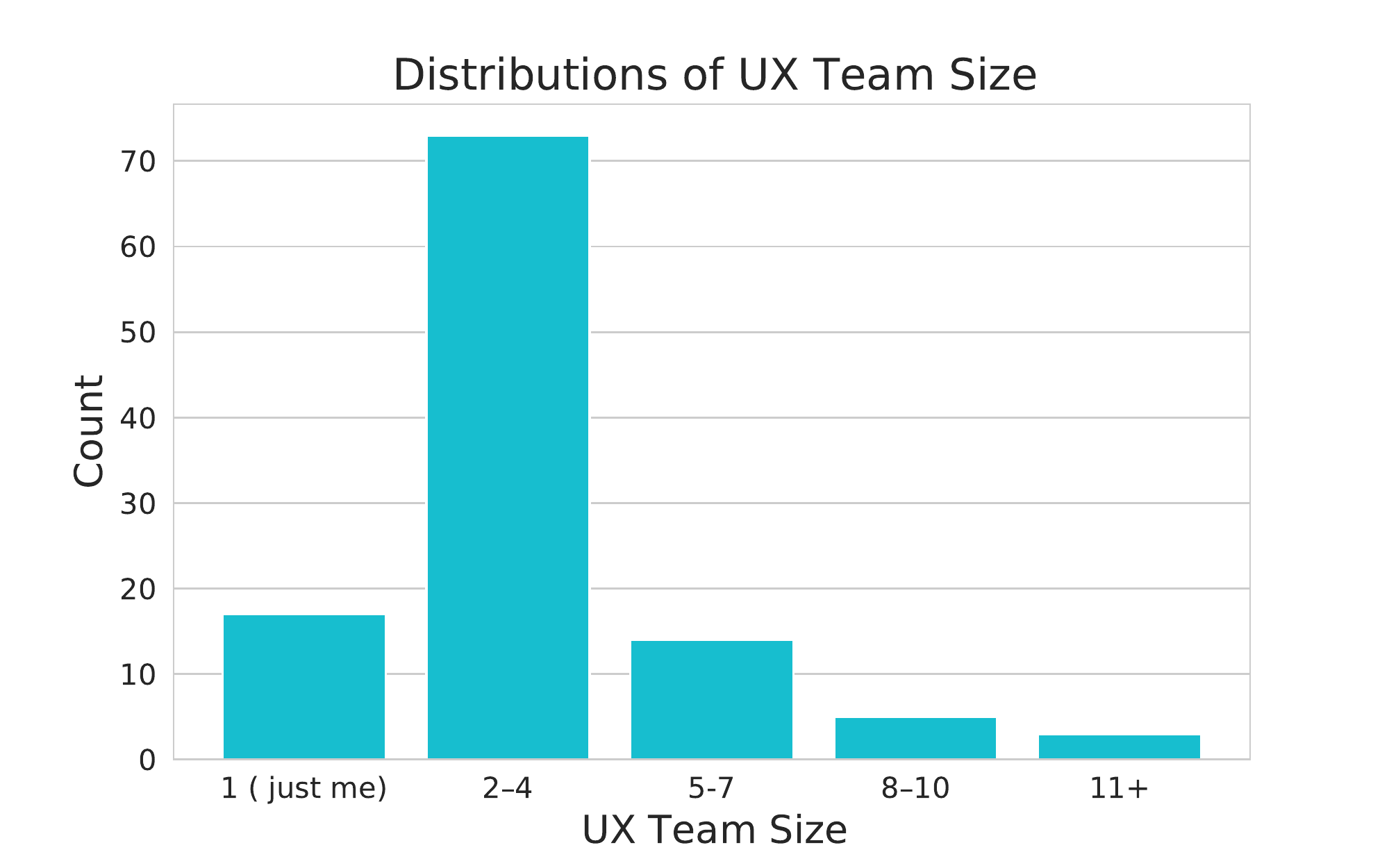}
        \caption{}
        \label{f:demo-3}
    \end{subfigure}
    \begin{subfigure}[h]{0.45\textwidth}
        \centering
        \includegraphics[width=\textwidth]{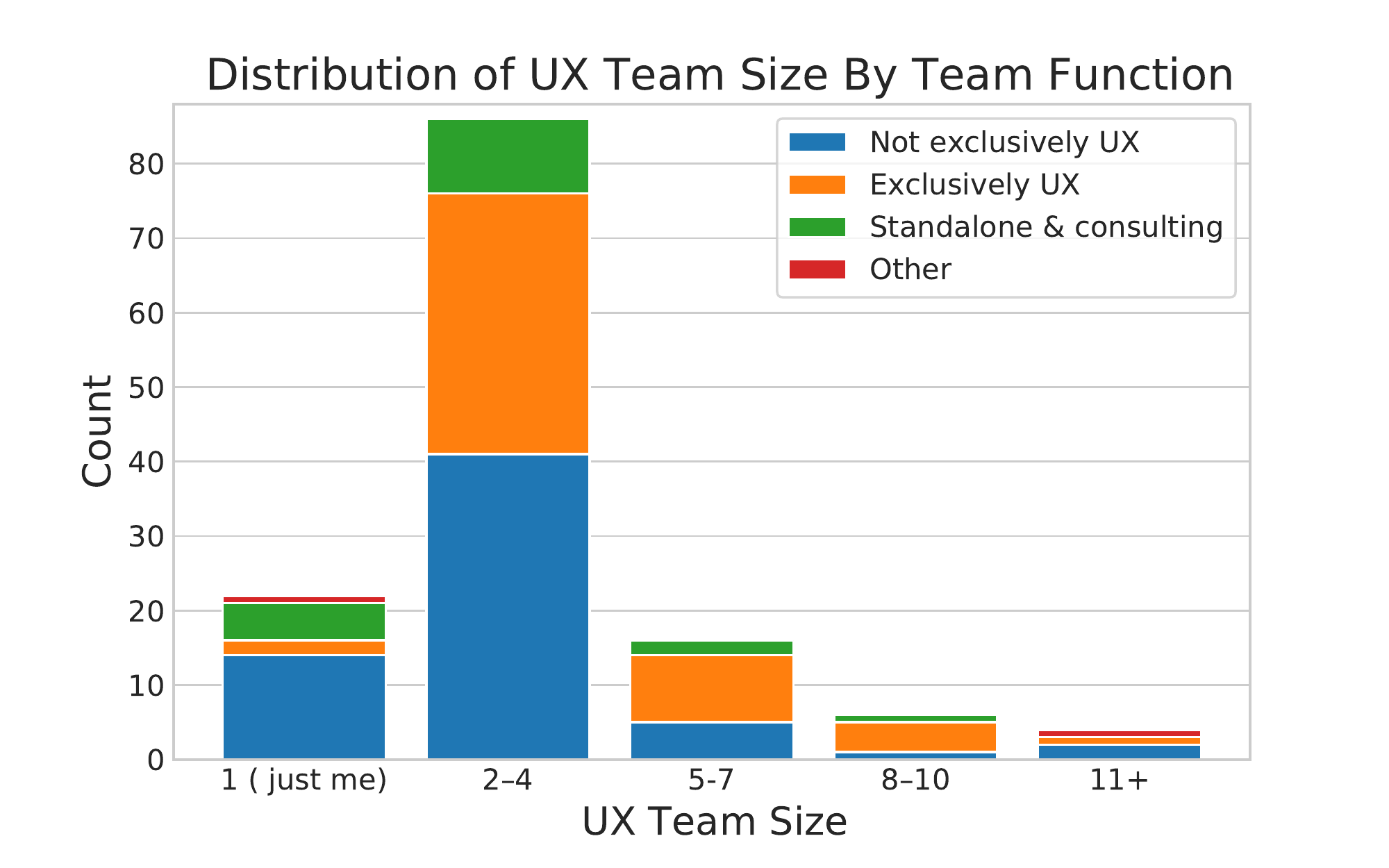}
        \caption{}
        \label{f:demo-4}
    \end{subfigure}
    \caption{Distributions of participants' (a) work experience, (b) employer size, (c) UX team size, and (d) UX team size by team function. Note that (c) and (d) do not share the same y-axis since participants can select more than one team function in (d).}
    \label{f:demo}
\end{figure*}

Out of our 114 participants, 59\% were female, 39\% were male, 1\% were non-binary, and 1\% preferred not to disclose. Most were early- to mid-career (see Fig. \ref{f:demo-1} for a detailed breakdown of years of experience in UX). Most were employed at large organizations with over 10,000 employees (see Fig. \ref{f:demo-2}). The UX teams in which our participants worked tended to be fairly small, with most teams consisting of 2–4 people (Fig. \ref{f:demo-3}). 38\% of participants' UX teams were situated within a larger team that did not exclusively do UX, 32\% were within a larger UX-only team, 10\% were standalone and consulted for non-UX teams, and the rest were some combination of at least two of the three. Participants were mostly UX designers (74\%); besides that, 16\% were UX researchers, 4\% were UX writers, 3\% were other (UX managerial roles), and the rest declined to answer. We had no participants who self-identified as UX Engineers.

We observed no significant correlations between participants' organization size and UX team size ($\chi ^2(12) = 9.47, p = 0.66$), but noticed that larger UX teams tended to be ones that focused exclusively on UX (Fig. \ref{f:demo-4}). Single-person teams were mostly situated within a non-UX team.

\subsection{Analysis}
We took both quantitative and qualitative approaches to analyze our survey data. For quantitative analysis, we performed statistical tests (e.g., Chi-squared test of independence, Friedman test) to determine if there were statistically significant associations between participant groups and/or stages in the design process. For well-structured free response questions, such as ones where we asked participants to list tools they used, we converted the responses into lists of categorical variables for statistical testing. We also qualitatively classified such responses into higher-level categories where applicable. For less-structured free responses, such as when we asked participants to share challenges in adopting design systems, we compiled all the responses to a particular prompt into one document. One author then took two passes through the data, first performing an open coding procedure to capture meaningful statements, and then organizing the results into broader themes through axial coding. 

\section{Findings}

With the survey data, we analyzed responses to questions relating to cross-functional collaboration across UX design stages, tool usage and collaboration, reuse with design systems, and transitioning design work to development.

\subsection{Collaboration Across Different UX Stages}
\label{s:collab-stage}
Professional UX work has necessitated cross-functional collaboration in organizational workflows. Indeed, all but 5 participants (96\%) indicated that they collaborated with others in their work. With this prevalence of collaboration in mind, we asked: \textbf{with whom do UXPs collaborate in each stage of the design process?} In addition to surveying the stages of the design process (Research, Synthesis, Low-Fidelity Design, High-Fidelity Design, and Communication) in which participants were involved, we collected participants' reported UX roles, UX team sizes, and team structures. 

\subsubsection{Product Managers and Designers as the Strongest Collaboration Partners}

\begin{figure*}
    \centering
    \begin{subfigure}[h]{1\textwidth}
        \centering
        \includegraphics[height=0.4\textwidth]{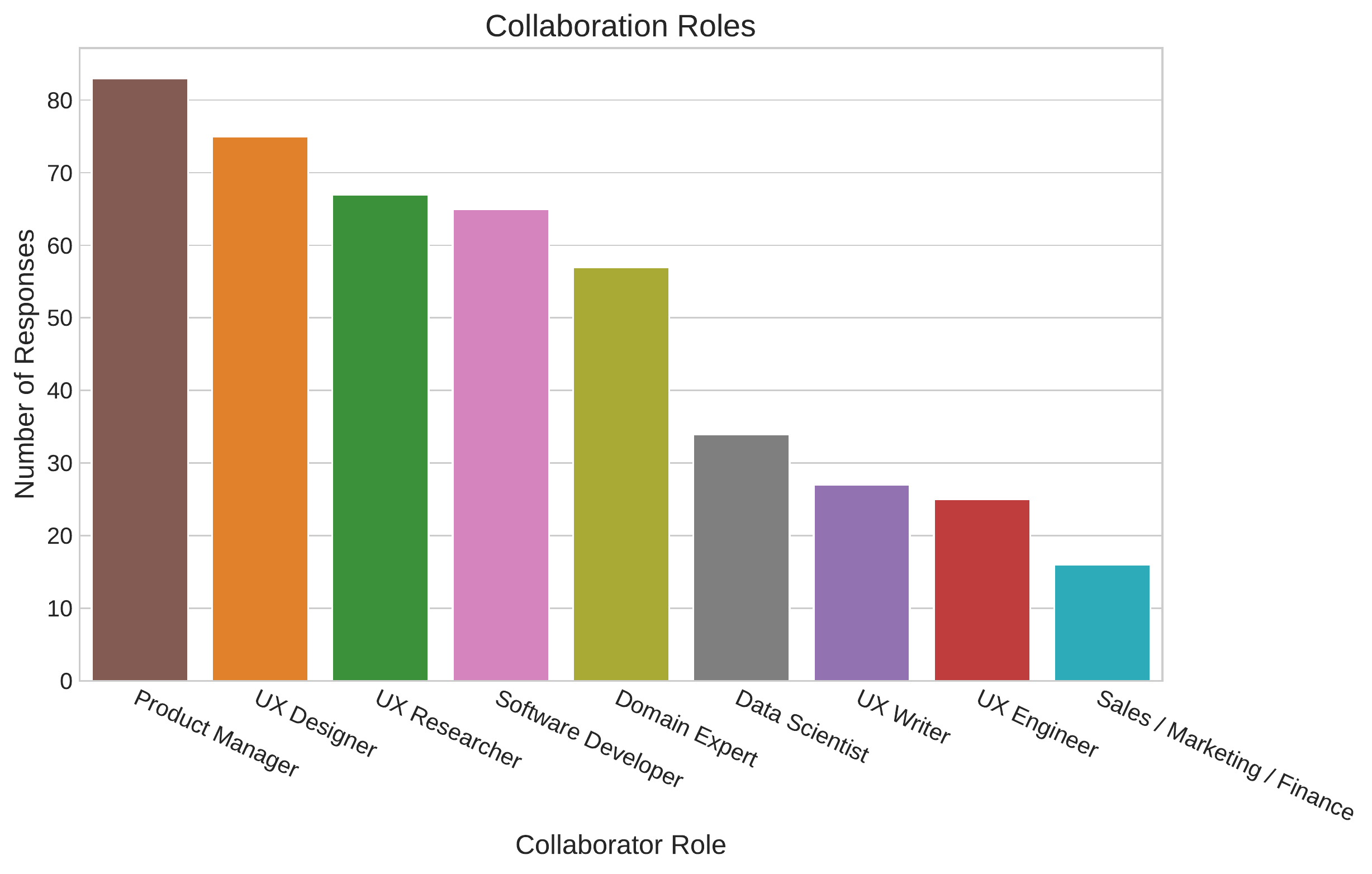}
        \caption{\add{How many participants reported working with each type of collaborator role at least once through each design phase ($n=112$). Overall, UXPs collaborated most with product managers, UX designers, UX researchers, and software engineers, in that order.}}
        \label{f:collab-overview-agg}
    \end{subfigure}
    \begin{subfigure}[h]{1\textwidth}
        \centering
        \includegraphics[height=0.6\textwidth]{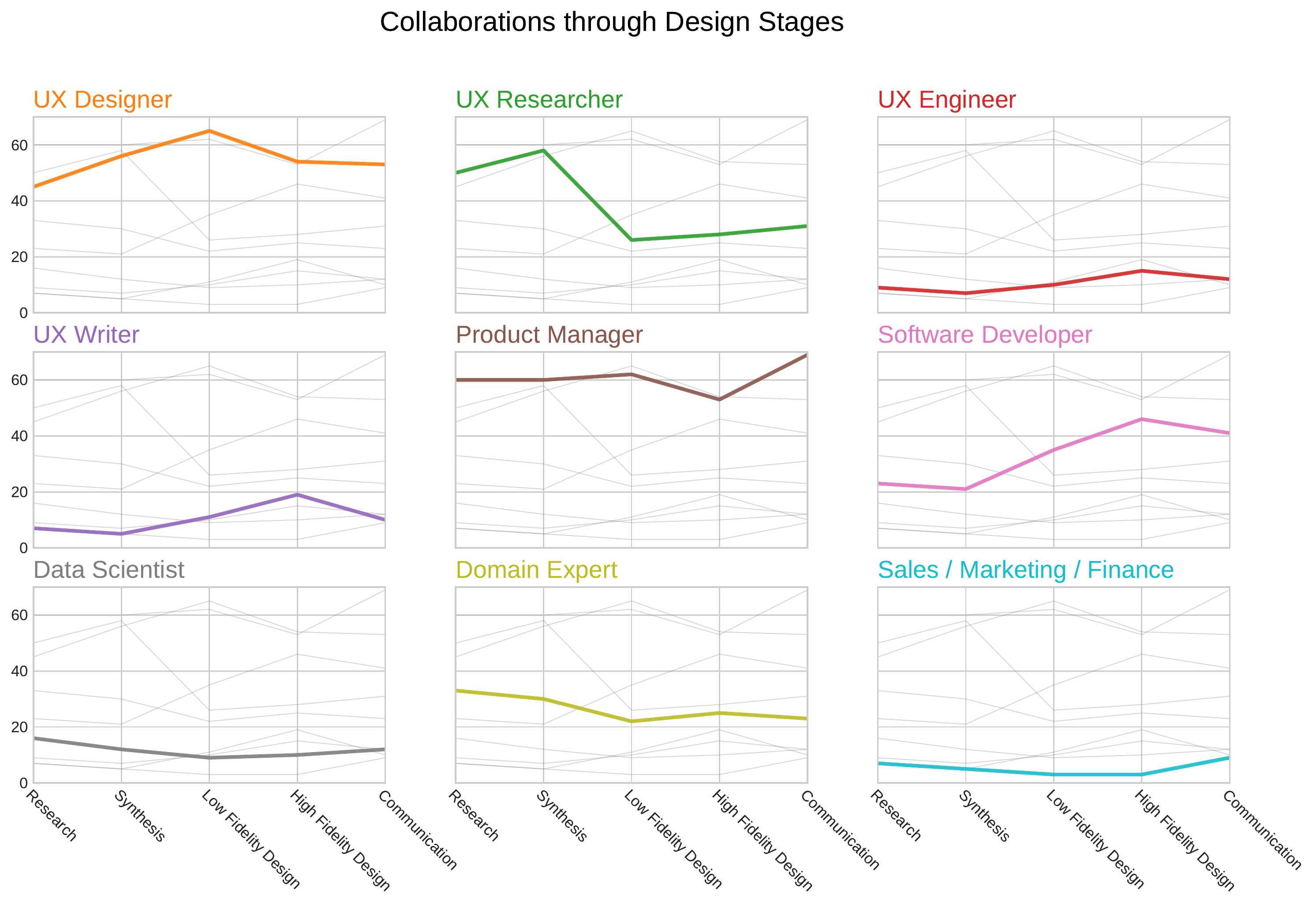}
        \caption{How many participants reported working with each type of collaborator role, through each design phase ($n=112$). Overall, collaborations with product managers and UX designers were consistently frequent throughout the design process, collaborations with UX researchers were more frequent in the early design stages than later stages, and collaborations with software developers were moderately more frequent in the later design stages than earlier stages.}
        \label{f:collab-overview-breakout}
    \end{subfigure}
    \caption{\add{Aggregated and disaggregated views of participants' collaboration. Roles are colour-coded between the two charts.}}
\end{figure*}

Participants who disclosed their collaboration patterns ($n=112$) generally reported working with UX designers and product managers the most (see Fig. \ref{f:collab-overview-agg}). Those who identified as designers worked with more UX researchers in the early design process stages (research and synthesis), with other designers more in the low- and high-fidelity design stages, and with product managers and software developers in the later design process stages (low/high-fidelity design and communication). Those who identified as researchers worked with both designers and product managers more in the earlier stages of design (research and synthesis). Noticeably, UXP interactions with software developers were usually limited to the UX designers; UX researchers did not collaborate with software developers often\add{ (see Fig. \ref{f:collab-overview-breakout}). Some collaboration patterns varied  across organization and team size---for example, larger companies and teams saw more collaborations with UX researchers, especially in the earlier design process stages, due to the presence of dedicated research teams within these organizations. Standalone UX consulting teams or single-person teams collaborated more frequently with domain experts throughout the design process.}



\begin{figure*}
    \centering
    \includegraphics[width=0.95\textwidth]{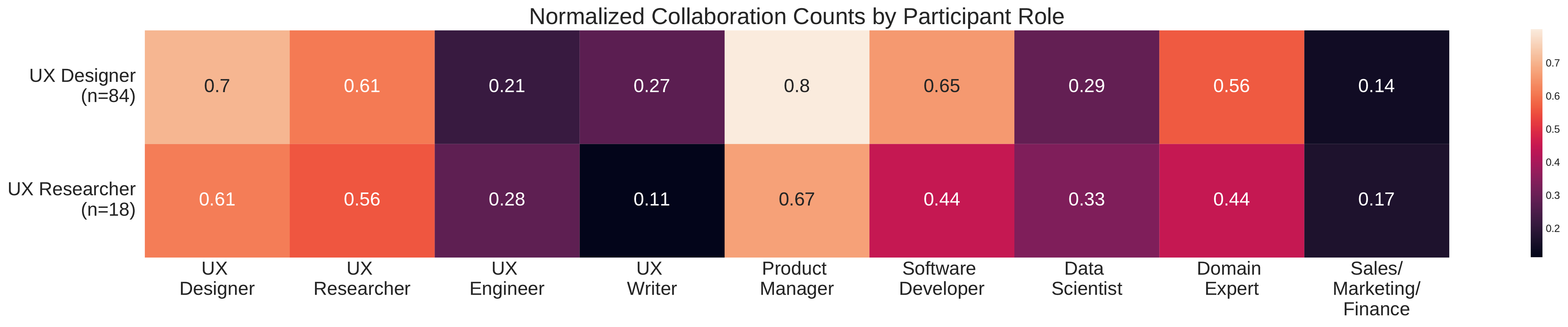}
    \caption{\add{Normalized counts of collaborators' roles (x-axis) broken out by participant role (y-axis). PMs, UX Designers, UX Researchers, and Software Developers are the most common roles that Designers and Researchers collaborate with. Only 12 participants identified as other roles; they are excluded from this figure.}}
    \label{f:collab-role}
\end{figure*}

\subsubsection{UX Practitioners Commonly Initiated Collaborations}
\label{s:collab-init}
We asked participants how their collaborations with non-UXPs were typically initiated. Out of those who responded ($n=108$), over half (63\%) initiated the collaboration because they or their team wanted others' input. As one participant explained, \textit{``I initiated because I wanted to give input. Non-UXers often forget to include UX. We include ourselves even when not invited.''} A substantially lower fraction (28\%) had their non-UX collaborators initiate the collaboration as the collaborators sought UX input. The rest experienced a combination of both. This imbalance in collaboration initiation is in some ways expected---previous work \cite{gray2015, thompson2010managing} has shown that embracing UX can be challenging for organizational cultures, and UXPs may need to go to extra lengths for their work to be recognized and understood. However, given the rapid proliferation of UX \cite{nielson2017} and establishment of UX teams in technology companies---to which many of our participants belonged---it was surprising that collaboration was still decidedly reliant on UXPs taking the initiative.

\subsubsection{Summary}
In general, UXPs appeared to have strong collaborations with product managers and UX designers throughout the entire design process, with some collaboration with UX researchers at the beginning of the process and with software developers later on. \remove{Larger teams and companies saw more collaborations with UX researchers, especially in the earlier design process stages, compared to smaller teams and companies. However, standalone consulting teams or single-person teams collaborated with product managers and domain experts more than with UX designers.} \add{Larger teams and companies saw more collaborations with UX researchers, while smaller ones collaborated more closely with domain experts.} Regardless of organization and team size, UXPs usually initiated these collaborations to gather others' input on their work.

\subsection{Tool Use and Collaboration}
\label{s:tools-methods}
In recent years, there has been increasing interest in collaborative and cloud-based UX tools \cite{2021-tools}. This led us to ask: \textbf{what tools are UXPs using throughout their workflows, and how are UXPs collaborating in these tools?} Through free-form text responses, we asked participants to list their toolset in each stage of the design process and describe any tool-based collaboration practices they engaged in. Some participants responded with conceptual tools (e.g., journey maps) in addition to software or physical tools; we ignored those as they were outside the scope of our ``tool'' definition in this paper. One author then turned the responses into tool lists for tools that were mentioned by more than one participant. Tools were tallied based on the number of participants who mentioned them (as an estimate of popularity) and iteratively organized into 10 high-level categories based on their functionality (see Table \ref{t:tools}). The author also open coded the collaboration practices described by participants. Below, we discuss our findings in tooling and tool-based collaboration in further detail.

\begin{table*}
\centering
    \begin{tabular}{p{5cm} p{9cm}}
    \toprule
    \textbf{Tool category} & \textbf{Tool (\# participants who mentioned this tool, >2 only)} \\
    \midrule 

    user research & UserTesting (5), SurveyMonkey (2) \\
    visual diagramming & FigJam (32), Miro (20), Mural (19), Google Jamboard (2) \\
    sketching & pen and paper (18), physical whiteboard (5), Balsamiq (2), Procreate on iPad (2)\\
    design & Figma (80), Adobe XD (11), Sketch (11), InVision (9), Adobe Illustrator (2), Axure (2), Flinto (2)\\
    design delivery & Adobe After Effects (4), ProtoPie (3), Zeplin (2)\\
    programming & frontend code (5), Swift Playgrounds (2)\\
    communication software \& artifacts & Microsoft Teams (15), Slack (8), Email (5), Outlook (4), videos (4), Google Meet (3), PDF (3), spreadsheets (3), calls (2), Zoom (2)\\
    document sharing \& editing & Google Docs (11), Excel (9), Word (7), Google Sheets (6), Google Suite (2), OneNote (2)\\
    project management & GitHub (4), Notion (3), Airtable (3), Quip (2)\\
    presentation & PowerPoint (30), Google Slides (10), Keynote (6)\\
    
    \bottomrule
    \end{tabular}
    \caption{Categories of tools used by UXPs in their workflow and the number of participants who mentioned using each tool. Note that \textbf{design} refers to tools that support end-to-end interface design from low-fidelity wireframes to high-fidelity prototypes, whereas \textbf{design delivery} tools take high-fidelity designs and prepare them for publication/testing/handoff.}
    \label{t:tools}
\end{table*}

\subsubsection{Figma as an All-In-One Tool}
\begin{figure*}
    \centering
    \begin{subfigure}[h]{0.33\textwidth}
        \centering
        \includegraphics[height=0.66\textwidth]{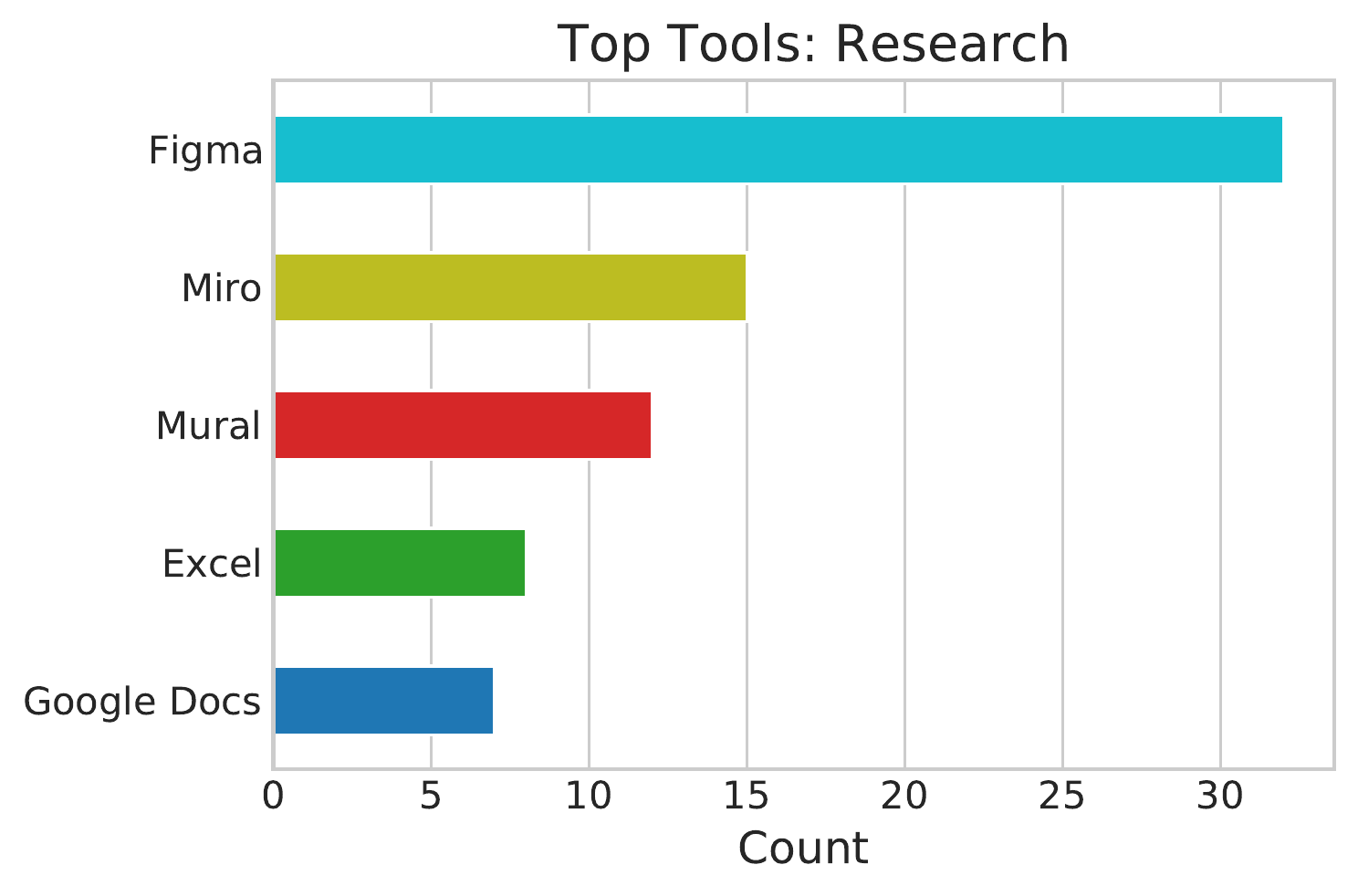}
        \caption{}
    \end{subfigure}
    \begin{subfigure}[h]{0.33\textwidth}
        \centering
        \includegraphics[height=0.66\textwidth]{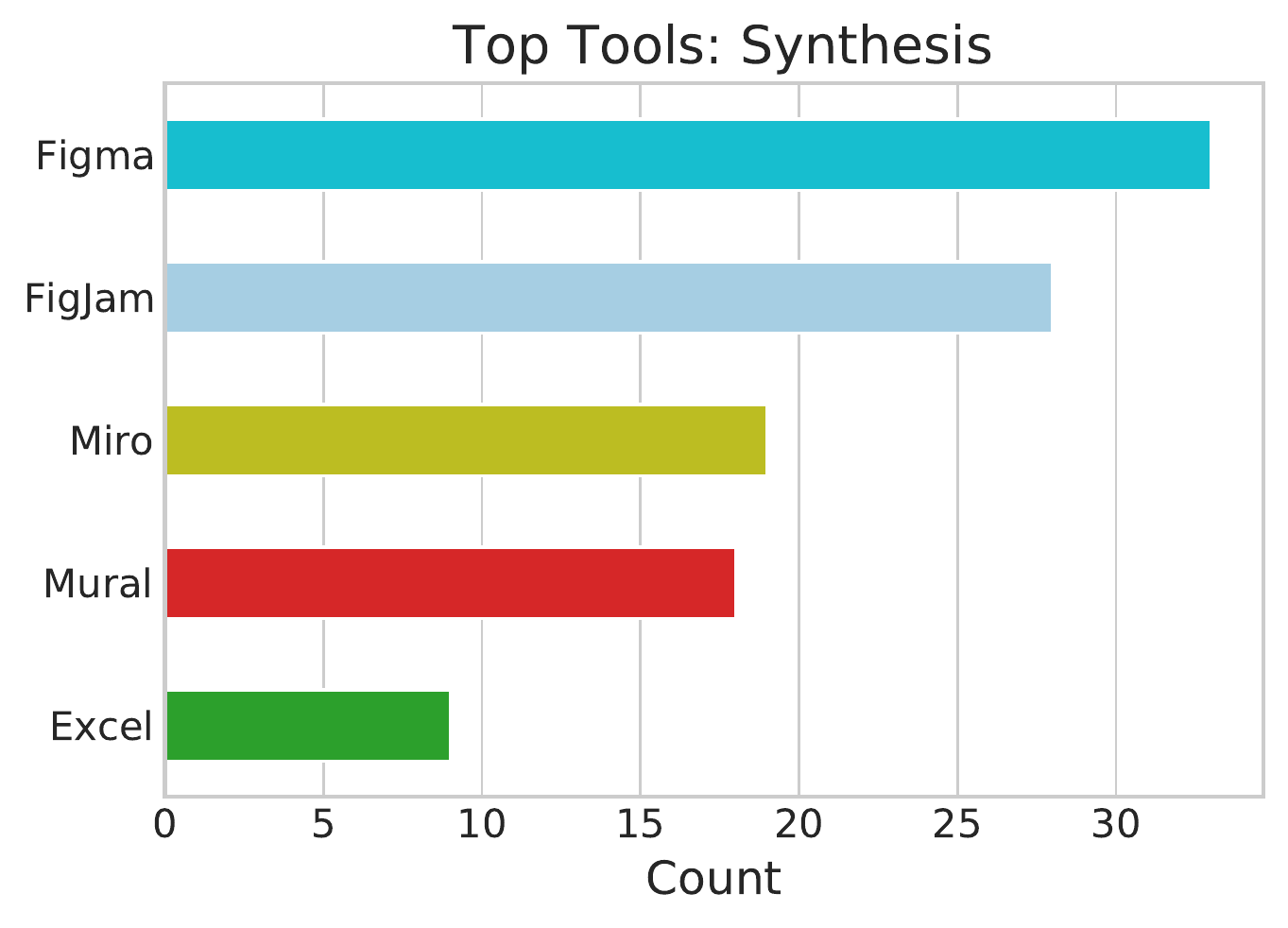}
        \caption{}
    \end{subfigure}
    \begin{subfigure}[h]{0.33\textwidth}
        \centering
        \includegraphics[height=0.66\textwidth]{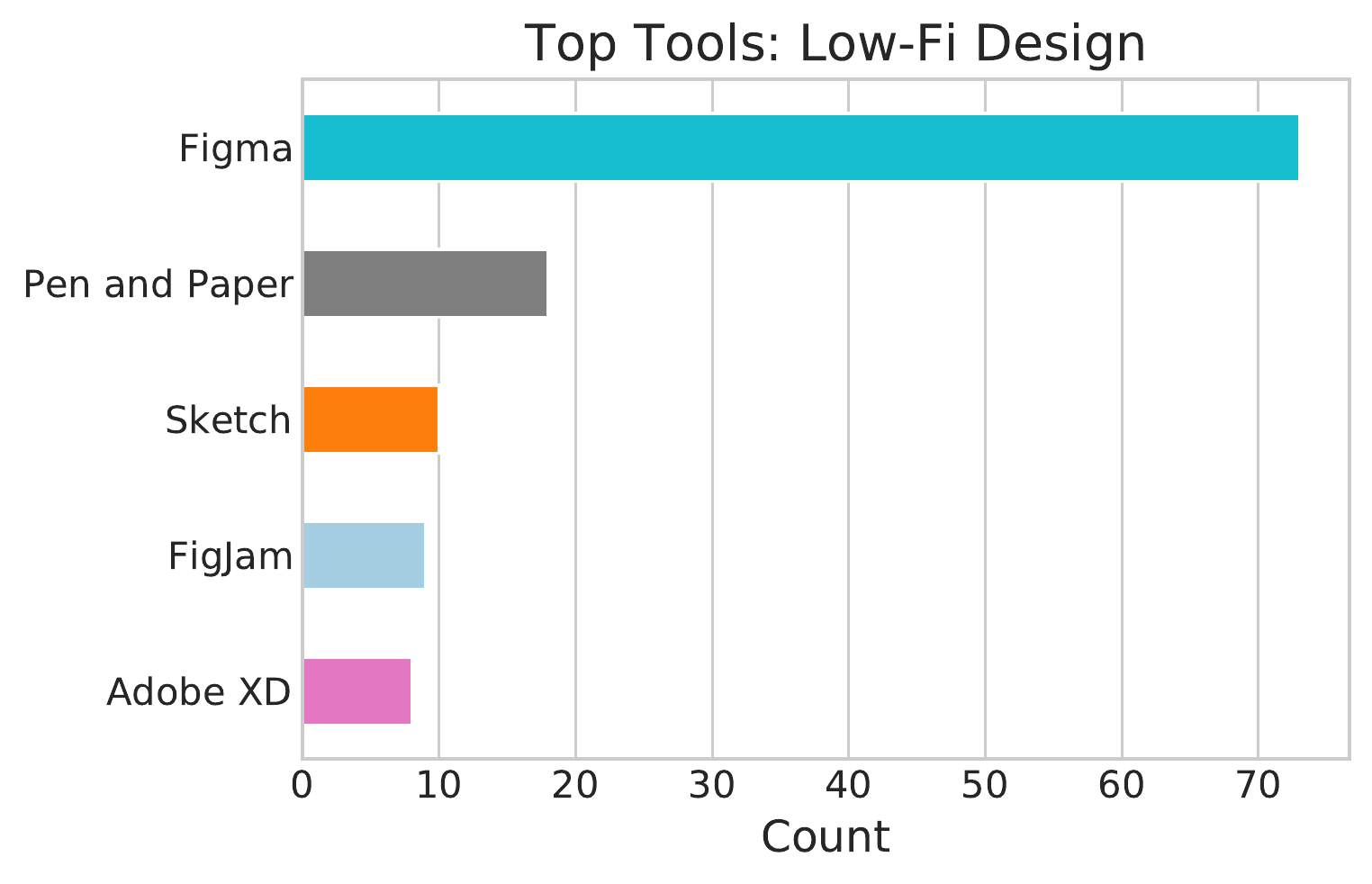}
        \caption{}
    \end{subfigure}
    \begin{subfigure}[h]{0.33\textwidth}
        \centering
        \includegraphics[height=0.66\textwidth]{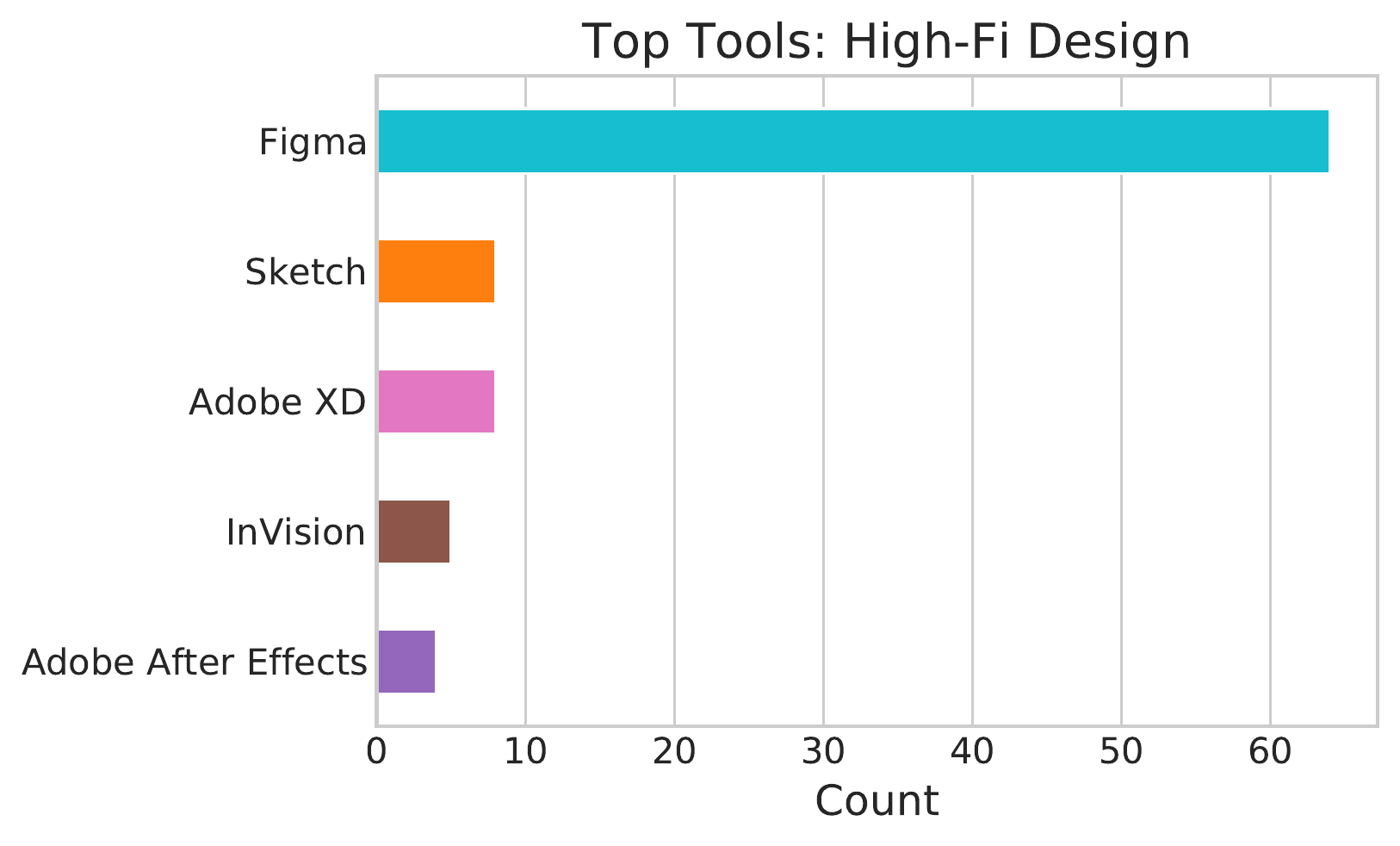}
        \caption{}
    \end{subfigure}
    \hspace{1em}
    \begin{subfigure}[h]{0.33\textwidth}
        \centering
        \includegraphics[height=0.66\textwidth]{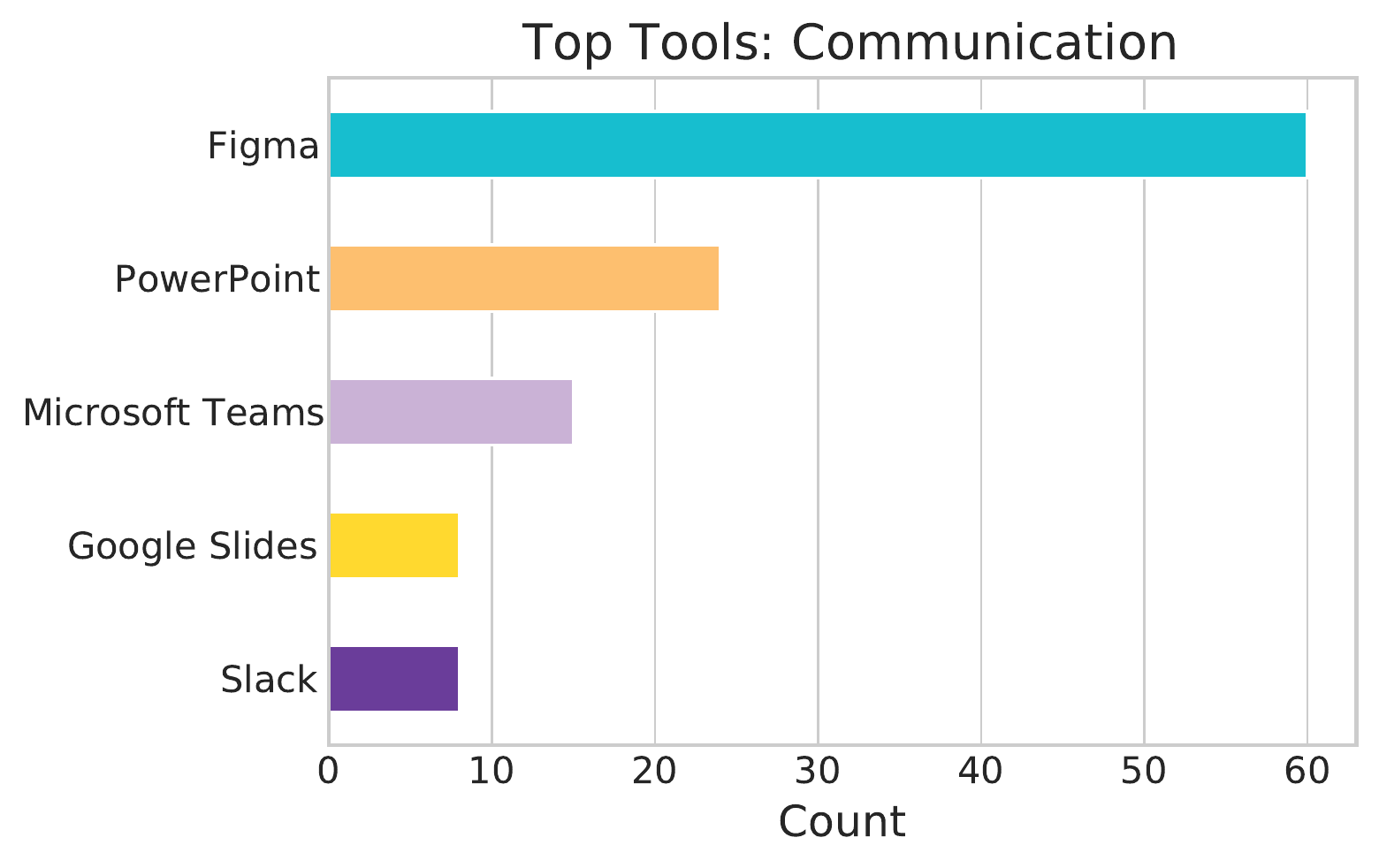}
        \caption{}
    \end{subfigure}
    \caption{Top 5 commonly used tools in each of the 5 UX stages and the number of participants who mentioned them.}
    \label{f:top-tools}
\end{figure*}

We examined differences in tool use across UX stages by using a Friedman test and found the differences to be highly significant ($Q = 33.20, p < 0.01$). The most common tools used for each stage are shown in Fig. \ref{f:top-tools}. 

The use of Figma was dominant across all 5 stages, aligning with similar findings from a 2021 survey on design tools \cite{2021-tools}. As one participant commented, \textit{``Figma has been an all-in-one tool, purely from the perspective that we can design, collaborate and update all within one file.''} Indeed, the Figma ecosystem (which includes FigJam) proved to be at least twice as popular as the second-place tool in all stages. This revealed the presence of an all-encompassing ``Swiss Army knife'' for UX workflows. Tools with similar properties also exist in workflows of other software-based domains, such as data science, but those domains do not necessarily exhibit the same concentration of tool popularity observed in UX \cite{zhang2020data}.

Not surprisingly, the related stages of Research and Synthesis had significant overlaps in tools: Figma, Miro, Mural, and Excel were all among the top 5 in each stage. Similarly, Figma, Sketch, and Adobe XD were all in the top 5 for both low-fidelity and high-fidelity design. All top 5 tools across the stages contained collaborative features or properties, with the exception of pen and paper during remote work. Most also contained social affordances for co-presence, such as multiple cursors and profile photos of other users if they are active in the tool.

\subsubsection{Synchronous and Asynchronous Collaboration in a Single Tool}
\label{s:collab-methods-tools}
In addition to asking participants about the tools they used, we also asked them how they collaborated and communicated in those tools. Working in \textit{shared, multiplayer canvases} stood out among responses---it was common for many collaborators to work and interact with each other from within one file. \remove{Teams may jump from one canvas to another throughout the design process, as one participant described: \textit{``shared documentation on Notion \& shared board on Mural \& shared prototype in Figma.''} }We \remove{also }observed two common roles emerging from workflows in shared canvases: author and editor. As one participant put it, \textit{``usually one person generate[s] the initial output, then others add on and/or comment on refinement.''} The author did not necessarily have to be a UXP, although that was often the case. Sometimes, software engineers and domain experts were invited to mock up their ideas in a collaborative design environment, and the UXP would refine and iterate on those ideas. This was valuable as participants collaborated in shared canvases both synchronously and asynchronously. 

\textbf{Synchronous collaboration.} \remove{A common practice for synchronous collaboration was to initiate a meeting and have everyone in the meeting also join the shared canvas to work, move artifacts around, and discuss. Some participants expressed a preference for this style of} \add{Some participants expressed a preference for synchronous} collaboration as their collaborators tended to not review their work asynchronously under the expectation of a future presentation: \textit{``some people go ahead and access the file on their own time but most people just wait for me to do a walkthrough presentation to review it and provide their feedback.''} Indeed, when working with other UXPs, \remove{synchronous design critiques---structured sessions where participants reviewed and gave constructive feedback on each other's work---were central. The } \add{the }verbal aspect of \add{synchronous} design critiques was seen as valuable: despite annotating on prototype screens and taking notes on the canvas and elsewhere, many participants still felt that much of their justification and documentation happened verbally. One participant mentioned that they recorded voiceovers walking through user flows to better explain their thought process. 

\textbf{Asynchronous collaboration.} Participants often used commenting features and virtual sticky notes to leave traces of information and feedback for others when collaborating asynchronously. Comments were mentioned as effective for working with those in another time zone, or to non-UX stakeholders: \textit{``Most non-designers are used to text-based feedback in Quip\footnote{Quip is a document and note writing tool: \url{https://quip.com}} so I’d write documents and have them comment.''} Commenting was distinct from the act of \textit{explaining} design decisions, which were often documented via text-based annotations directly on or beside their designs. Participants also documented their decisions by taking advantage of the version-tracked design files to show the progression of mockups over time, overlaid with research insights and/or technical constraints from developers. These \textit{layered artifacts}\footnote{We borrow this term from Schmidt and Wagner \cite{schmidt2002coordinative}, who used it to refer to annotated CAD drawings with similar collaborative properties observed in architectural practice.} helped participants frequently reference user or academic research, and actively share lower-fidelity work such as wireframes and user journeys.

\subsubsection{Presentation of UX Work Involved Different Techniques and Environments} 

When presenting UX work, participants often left the shared canvas for presentation software, such as PowerPoint, or entered a special mode of the workspace, such as Figma's prototype mode.\footnote{The prototype mode is a separate interactive environment that simulates how end users may interact with designs \cite{figma-prototype-mode}.} This was due to some higher-level leadership not being familiar with Figma, as well as the ability to structure a better narrative in a slide deck or via prototype mode after the bulk of the UX work was complete.

Participants varied their presentation strategy slightly when showcasing work to non-UX collaborators compared to fellow UXPs. For one, presentations to non-UXPs tended to be more visual and higher-fidelity: instead of sharing drafts of wireframes, participants shared links to and presented clickable prototypes. Sharing an entire design file (common among UXPs) was less common; typical artifacts for communicating decisions were PDF documents, videos, slide decks, and articles. There was also a heavy emphasis on storytelling for non-UXPs: participants told success stories from clients, walked through the context and history behind the steps used to land on a certain decision, and created executive summaries, data visualizations, and other visual imagery to support their work. According to one participant, the storytelling was typically done \textit{``in layman's terms and include[d] lots of high-level overview slides.''} Some participants also mentioned that they would establish clear links between their decisions to business benefits and outcomes, with at least one participant explicitly mentioning that they discussed their decisions' financial opportunities.

\subsubsection{Summary}
In Section \ref{s:collab-stage}, we saw that UXPs actively engaged in collaboration with various roles throughout the design process. Here, we show that collaborative, multiplayer tools hosting shared files to be used across multiple workflow stages were widely used among UXPs and their collaborators. UXPs employed a variety of strategies for synchronously and asynchronously collaborating in these shared environments, but the strategies can differ based on whether they were communicating with other UXPs or non-UXPs. 

\subsection{Collaboration \remove{with Technical Stakeholders} \add{During Handoff}}
\label{s:handoff}
In the words of one participant, \textit{``the real world challenge starts when it comes to implementation.''} We asked participants about their current practices and challenges when they handed off their UX work to software developers for implementation, paying close attention to the scenarios where the design was different from implementation. We refer to this phenomenon as \textit{design-development divergence} as the original goal of alignment between design and code is often discarded due to procedural breakdowns over time \cite{maudet2017breakdowns}. For participants who had experience in designing machine learning (ML)-powered user experiences, we asked about their collaboration practices and challenges \add{with ML practitioners (e.g. data scientists, ML engineers). We did so given recent interest in enhancing communication and handoff between UXPs and ML practitioners (who are distinct from software developers \cite{amersio2019swe}), which has been noted as uniquely difficult in prior literature} \remove{as prior literature in UX design ML has noted and proposed solutions for the uniquely difficult challenges in this area} \cite{yang2020difficult, subramonyam2021protoai, subramonyam2021towards}. Unlike our findings in Section \ref{s:tools-methods}, this section highlights some challenges and desiderata of \textit{sequential} collaboration---where UXPs and developers or data scientists work one after the other in different domains---rather than \textit{parallel} collaboration---where UXPs and developers work together within a single stage of the design process.

\subsubsection{(Over)communication During Handoff}
\begin{figure*}
    \centering
    \includegraphics[width=0.7\textwidth]{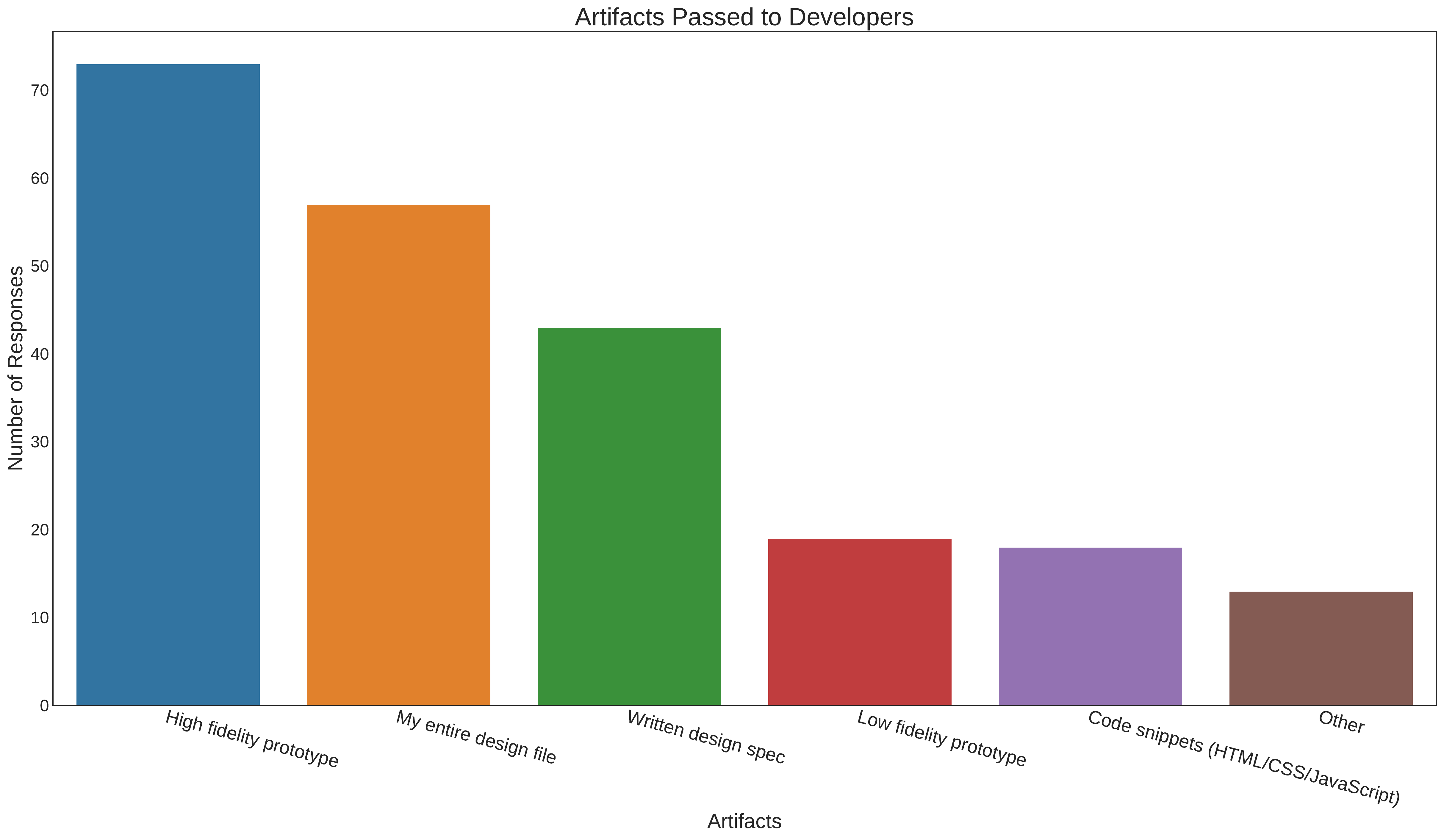}
    \caption{Distribution of common artifacts shared by UXPs with developers as part of the handoff process ($n=88$). }
    \label{f:artifacts-distr}
\end{figure*}

Out of the 96 participants who participated in handoff, the vast majority (92\%) shared their work with developers for handoff. Participants used a range of artifacts to share their designs with developers, including high-fidelity prototypes, the entire design file, written design specs,\footnote{Design specifications (specs) are detailed documents prepared by UXPs outlining UI design choices (colours, character styles, measurements), as well as interactions (user flows, component behaviours, and UI functionality) 
\cite{design-specs}.} and CSS code snippets generated by design tools. High-fidelity prototypes were the most commonly shared artifact, followed by the entire design file (see Fig. \ref{f:artifacts-distr}). Most (80\%) shared more than one artifact; the most common combination of artifacts was the design spec, high-fidelity prototype, and the entire design file (17\%), followed by the same combination but without the design spec (11\%). 

Sharing the entire design file was convenient in collaborative tools, but participants pointed out pitfalls with \textit{too much transparency}. For instance, the lack of a constrained view where one can control which specific screens collaborators can see may be troublesome: \textit{``With Figma you can point someone to a specific frame but they can just zoom out and access every other page or frame in the entire file which often leads to confusion and frustration.''} Some participants mitigated this issue by clearly demarcating a region in the file for the final product, with watermarks and naming conventions that indicate work-in-progress designs. In general, participants relied heavily on attention-grabbing text annotations (or even paragraphs of writing) in the design file itself to explain UI behaviours. Unlike providing UX feedback (Section \ref{s:collab-methods-tools}), these annotations are typically \textit{not} written via the tools' commenting features as developers can forget to view the file on ``comment mode'': \textit{``I will write comments in hot pink on the screen so it is obvious that it is not part of the UI. I do this because you have to go to comment mode in Figma to see comments, [which is] easy to miss.''} In addition to annotations, participants also ``redlined'' parts of their design---adding red lines and text to mark up spacing and sizing of design elements. 

These text-based strategies alone do not solve some broader procedural challenges, such as establishing alignment while implementation happens. As one participant noted, \textit{``Design/UX work doesn't change as much as a developer's work, so you have to make sure you have ample time to work and still align.''} Indeed, even after handoff, participants frequently made themselves available to answer questions and participated in ``bug bash'' sessions where they inspected the implementation with developers and logged visual ``bugs'' that did not align with their original design specifications. To organize and streamline the design-to-code process, one participant even broke down the screens they designed into smaller implementation tasks which they then recorded as GitHub tickets, as that allowed them to converse with developers on common ground. 

\subsubsection{Encountering and Handling Design-Development Divergence}
\begin{figure*}
    \centering
    \includegraphics[width=0.75\textwidth]{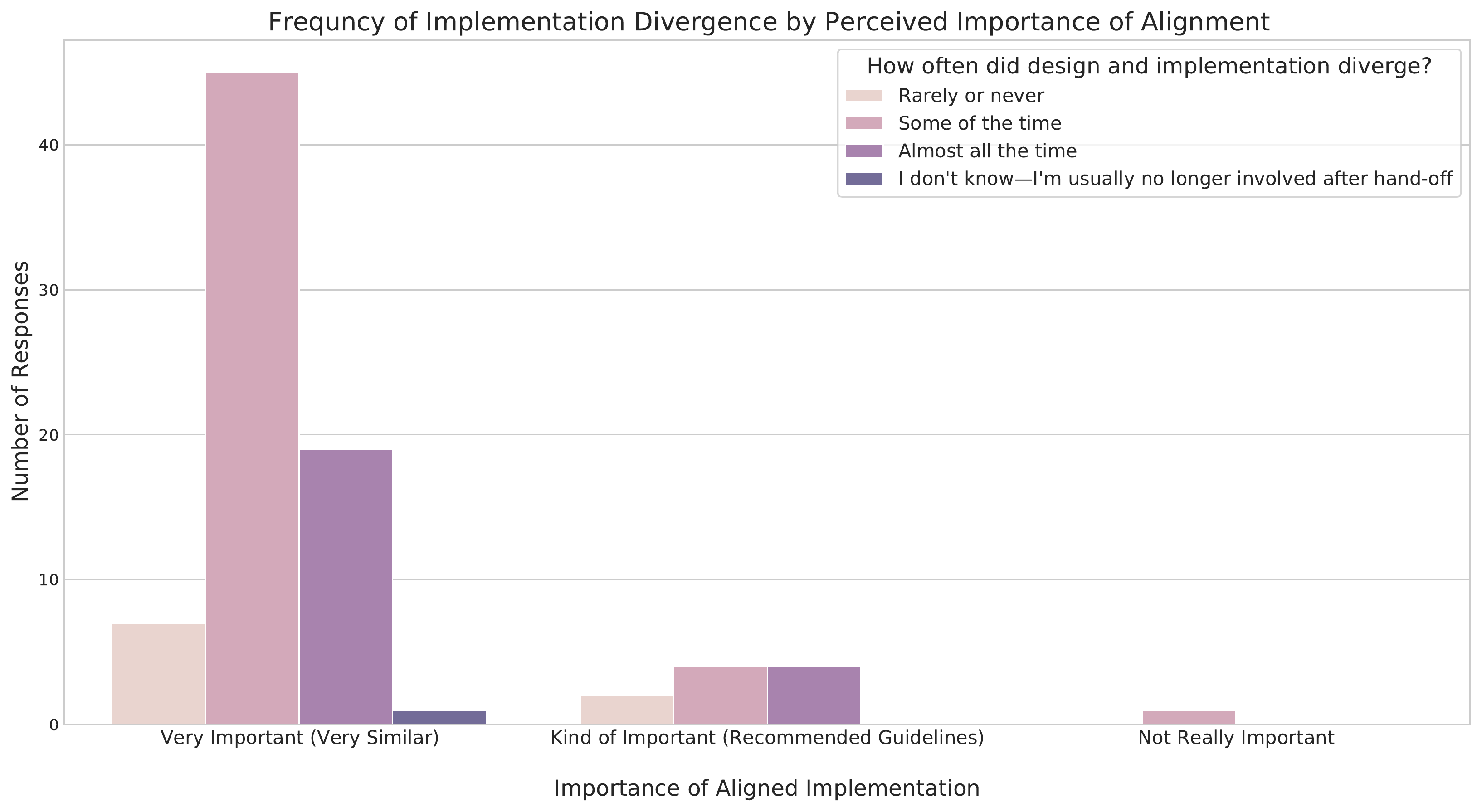}
    \caption{\add{Frequency of design/implementation divergence by perceived importance of alignment. Most respondents considered similarity important but saw more divergences than not.}}
    \label{f:divergence-importance}
\end{figure*}

Participants considered it to be very important that developers' implementation closely match the look and feel of their designs (86\% of $n=87$ respondents). A much smaller fraction (13\%) expected their designs to be treated as recommended guidelines for implementation, while only one respondent thought developers could freely follow or reject their designs at will. Despite this, expectations did not align with reality: 57\% sometimes encountered design-development divergence, while 26\% encountered it almost all of the time \add{(see Fig. \ref{f:divergence-importance})}. This quantifies some results of previous work that investigated why design breakdowns can easily occur at design-development boundaries \cite{maudet2017breakdowns}.

More than half (59\%) reported that once divergence happened, design changes were necessary to bring their work closer to implementation. This was surprising as it points to implementation being the ``source of truth'' instead of an intermediate representation of design and implementation, as found in prior work \cite{maudet2017breakdowns, leiva2019enact}. Should this be how it is? Our participants appeared divided on whether design should conform to implementation, vice versa, or meeting somewhere in the middle. Many (41\%) ``caught up'' with the implementation by editing or even re-doing some of their designs. A lesser fraction (10\%) was adamant about the ``proper'' implementation of their designs and gave feedback to developers to better align with designs. 
A few participants acknowledged that designs can sometimes be technically infeasible---developers may run into API or environment constraints only after starting implementation. Direct communication with developers seemed to be the most effective way to identify the problem's root cause. One participant stated that upon deliberation, what was thought to be a technically infeasible design simply needed additional scenarios that developers encountered during implementation. Other courses of action after encountering divergence included continuing to use original, outdated designs but keeping in mind that they are different from the implementation (14\%), and discarding the original designs altogether to design off of screenshots of the implementation instead (10\%). 

\subsubsection{Knowledge and Process Gaps for  \remove{Machine Learning Collaboration} \add{Handoffs Involving Machine Learning}}

58 participants recounted their experiences on an ML-enabled project. \add{Commonly mentioned ML technologies include recommender systems (50\%), natural language processing---e.g. personal assistants, speech-to-text---(48\%), computer vision---e.g. object detection, photo tagging---(40\%), and spam \& malware filtering (9\%). Unlike handoffs in non-ML projects, where UX work preceded technical implementation, we observed that ML handoffs often followed a \textit{reversed} workflow where the ML model is implemented prior to the start of any UX work. Consequently, it was essential for UXPs to familiarize themselves with model capabilities to understand the feasible design space.} \remove{Most (74\%) learned about model capabilities indirectly through communicating with those in ML roles, such as data scientists and researchers. Frequent meetings with ML practitioners, watching them present their work about the models, and asking for demos were common practices to attempt to understand what models can and cannot do. This practice also commonly surfaced in previous work \cite{subramonyam2021towards, yang2018}.}\add{Most (74\%) learned about what models can and cannot do indirectly through frequent meetings with ML practitioners, watching them present their work about the models, and asking for demos, agreeing with prior work \cite{subramonyam2021towards, yang2018}.} A smaller fraction of participants (16\%) interacted directly with the model to learn more about it, feeding it test cases in sandbox testing environments as well as viewing common metrics for model evaluation (e.g., F1 score, ROC AUC). The remaining 10\% were not aware of model capabilities at all.

We saw two categories of challenges when UXPs collaborate with ML practitioners: \textit{knowledge} challenges and \textit{process} challenges. Knowledge challenges included technical jargon, which was often not adequately explained, and the inability to understand technical constraints of the model. However, it was more frequent for participants to mention \textit{process} challenges, an example of which is tricky timing:

\begin{quote}
    \textit{``A lot of times we want to test a concept with real data, but the models we're testing with aren't production-ready yet, or researchers have moved onto a different problem, so there can be kind of a lag for getting data, or getting feedback on how a model would theoretically work before we start usability.''}
\end{quote}

Participants also implied that their collaborations also posed a one-way burden on ML practitioners, needing to ``bug'' them to ask them questions and having them \textit{``explain [concepts] to me like I'm five.''} To mitigate this burden, participants adopted the strategy of orchestrating collaborations \textit{``in a way that it takes minimal effort for [collaborators]''} to share knowledge. Despite this, some noted that ML practitioners were quite excited about design implementations and were very interested in having a UX perspective, but pointed out the common pitfall of \textit{``seeing the design discipline as a service, not a partner''} \add{when expecting handoff artifacts from UXPs.} To ameliorate these challenges, participants wished for a way to ``translate'' technical requirements to more understandable UX contexts, such as user scenarios. Another popular suggestion was an interactive environment in which they can train their own variations of models to enhance their behind-the-scenes understanding. \add{For instance, one participant wanted a \textit{``plugin where you could enter in fake training data to simulate a prediction. Example: For an NLP [...] I might have 20-30 canned phrases and each would have a preordained result.''}} 

We were surprised that a few participants (7\%) did not believe \remove{that } designing ML interfaces was noticeably different than non-ML ones\add{, a sentiment not obvious in prior work.} \remove{: as} \add{As} one participant noted, \textit{``ML / AI [is] not some special, wild wild west.''} A common reasoning for this is that UX work primarily resides in the front-end interface layer, while AI is an abstracted back-end. This is perhaps best summarized by the following quote, the sentiments of which have been echoed by others:

\begin{quote}
    \textit{``For me, ML/AI are back-end tech, almost black-box in that I just need to know user inputs, behaviours, and outputs; how those are generated aren't essential.''}
\end{quote}

Divergent perspectives on whether designing with ML was uniquely challenging can perhaps be attributed to variance in the nature of projects, as one participant hinted: \textit{``\textbf{in this particular effort} the particular details of our ML model did not impact design decisions that much.''} ML projects may indeed fall into various levels of AI design complexity, as taxonomized by Yang et al. \cite{yang2020difficult}---ML systems with fixed capabilities and few possible outputs may require fewer specialized processes and tools than ones with constantly evolving capabilities and seemingly infinite outputs.

\subsubsection{Summary} Despite collaborative features and handoff guides being offered in UX tools \cite{figma-handoff}, UXPs still encountered persistent challenges similar to those outlined in previous work when handing designs off to developers. \add{Handoffs to ML practitioners were often procedurally distinct, almost reversed, compared to handoffs to non-ML developers. Still,} \remove{Similarly,}when designing for ML-enabled interfaces, many encountered knowledge and procedural challenges, expressing interest in new tools and shifts in collaboration styles. 

\subsection{Design Systems and Reuse}

In UX work, design systems are used to standardize and document design work for reuse, especially across larger organizations \cite{churchill2019scaling}. UX tools have made design system management more accessible through specialized features for storing, organizing, and updating contents of the system \cite{figma-plugins, xd-plugins, sketch-plugins}. We wanted to better understand the effectiveness of collaboration and reuse through design systems, so we explored our research question: \textbf{what are UXPs' current expectations and practices around reusing design components and managing design systems in collaborative UX tools?}

\subsubsection{Expectations Around Reusing Designs}


Most participants reused designs from other UXPs in their company (71\% of $n=95$), as opposed to designs from outside their company. This was especially true for large companies (>10,000 people), possibly due to confidentiality and higher maturity in the internal design system. Smaller companies had higher proportions of UXPs who did not use designs from others (36\% for companies of fewer than 100 people, 40\% for companies of 101-1000 people). For solo teams, 50\% of respondents did not use designs from others, a much higher proportion than for any other team size. Additionally, most UXPs expected their work to be reused (81.5\%), across all company and team sizes. This expectation aligns with the high response rate we saw for UXPs using assets from design systems.  


\subsubsection{Challenges in Design System Management}
To better understand obstacles to working design systems in collaborative tools, one author qualitatively categorized open responses detailing challenges in using and adopting design systems with collaborators. The most common responses were due to the design system not being able to scale to diverse collaborative teams (22\% of responses), followed by lack of communication and ambiguity in ownership (20\%). Refer to Table \ref{t:design-sys-challenges} for all categories of challenges and associated number of responses.

\begin{table*}
\centering
    \begin{tabular}{p{10cm} p{4cm}}
    \toprule
    \textbf{Reason} & \textbf{Number of Responses} \\
    \midrule 

    Collaborative team discrepancies with nonscalable design system & 13 \\
    Lack of knowledge communication with ownership ambiguity & 12 \\
    Design system changing too often & 8\\ Mismatch between design and code libraries/models & 7\\
    Lack of documentation or awareness of systems & 6\\
    Insufficient resource budgeting and efficiency & 6 \\
    None or N/A & 7\\
    
    \bottomrule
    \end{tabular}
    \caption{Categories of responses to the question "What are some challenges you observed in getting design systems adopted widely by your collaborators, if any?", with the most cited reason being design systems failing to scale to collaborative teams.}
    \label{t:design-sys-challenges}
\end{table*}

Based on these results, maintenance and update of design systems to serve the diverse needs of multiple user teams seemed to be the most difficult challenge for most UXPs. As one participant stated: \textit{``It's the classic challenge of trying to design something that works for everyone while still allowing the freedom of a given project to make a design optimized for their given needs.''} Communication---in this case, between design system creators and users---appeared to be the next most common challenge. With no clear ownership or guidance on who should update the design system and how to enforce standardization, UXPs struggled to find design systems useful. One participant mentioned: \textit{``We all use [a design system] if it's there, but since nobody has ownership, it may fail to have new assets or stylings.''} \add{Another pointed out that it can be ambiguous when \textit{``deciding what elements need to go into [the design system], and what don't,''} particularly for new designers who are just becoming familiar with the system. Finally, participants expressed frustration at the lack of support to transfer design systems between tools. One complained: \textit{``design systems are often arbitrarily linked to one program like Figma, which prohibits wide use,''} while another explained that every time their team upgraded their tools, they needed to recreate their design system, which prevented the system from scaling to desired levels. Furthermore, this opened up opportunities for versioning issues in design systems, which one participant described when their team used slider components: \textit{``we ended up using 3 different styles of sliders---ones from the `old' design system, ones from the `new' design system, or custom made ones.''}}

\subsubsection{Summary} Component reuse in collaborative tools was prevalent, particularly in larger organizations; however, new challenges emerged in managing design systems for structured reuse. Although the rigidity of design systems may be a reason for their adoption in the first place, it was also what makes them difficult to adapt to diverse use cases and accommodate more contributors. Some current practices \add{and inter-tool support} for collaborative design system management also appeared hazy and underdeveloped.

\section{Discussion}
\begin{figure*}
    \centering
    \includegraphics[width=1\textwidth]{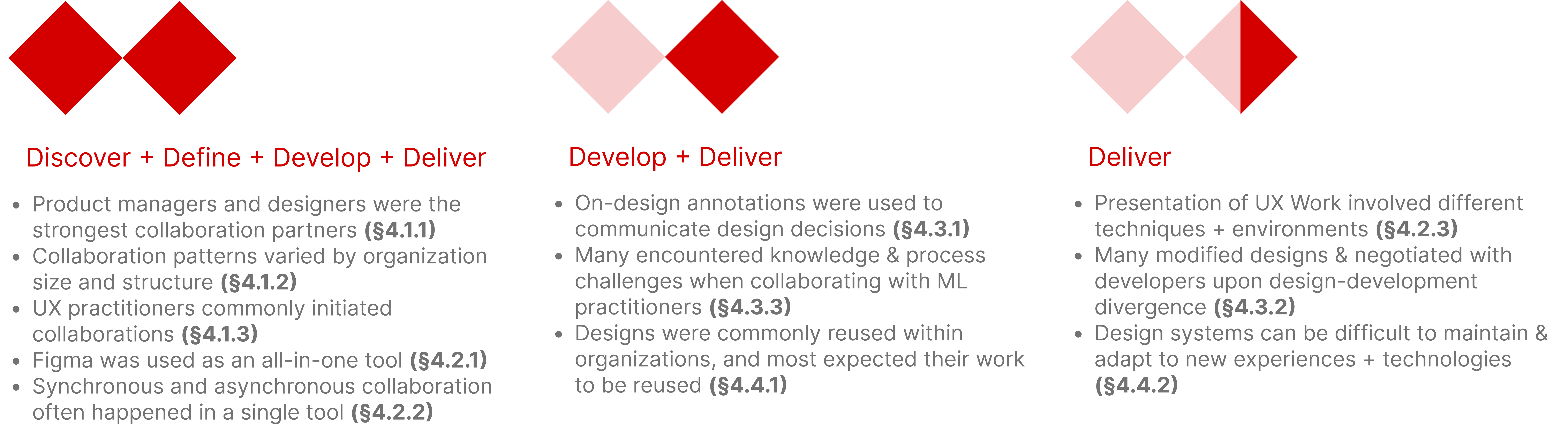}
    \caption{\add{Our main findings mapped onto the double diamond design process. Most of our findings span more than one stage. }}
    \label{f:f-dd}
\end{figure*}

Throughout our survey, we see that UXPs leveraged collaborative practices and tools throughout the design process to enhance their and others' work. We took a deeper dive into specific challenges outlined by previous work---namely designing with ML \cite{yang2018, yang2020difficult, subramonyam2021protoai}, developer handoff \cite{maudet2017breakdowns, leiva2019enact, walny2020handoff}, and design system management \cite{churchill2019scaling, moore2020systems}---and found those challenges to persist in collaborative environments. Furthermore, UXPs faced new challenges such as exposing extraneous information that confuses non-UX collaborators and writing attention-grabbing annotations to explain their work. \add{We map our findings onto relevant stage(s) of the double diamond given its importance in design practice (see Fig. \ref{f:f-dd}). Interestingly, findings that do not span the entire process lie only in the second diamond. This indicates that more collaborative challenges arise after the formulation of the design problem rather than before, but we note this may also be a bias due to our sample containing considerably more designers (who typically work around the second diamond) than researchers (who typically work around the first).} We unpack \remove{these} \add{our} findings and \remove{their} \add{corresponding} implications below. 

\subsection{UX Tools Act as a Hub, But Not Necessarily Practitioners}
\add{In RQ1, we asked, ``with whom do UXPs collaborate at each stage of the design process?''} As delineated in Section \ref{s:collab-methods-tools}, UXPs collaborate extensively in their tools throughout different stages of the design process. In particular, UXPs tended to invite collaborators (e.g., product managers, software developers) into tools typically considered to be for UX (e.g., Figma). Collaborators then ``jammed'' together---interactively brainstormed, created, and edited artifacts---synchronously and asynchronously in structured ways through assigned roles of authorship and editorship. \add{This tendency sheds light on RQ2: ``what tools and tool-based strategies are UXPs using throughout their workflows?''} 
\remove{This tendency suggests} \add{Our findings here suggest} that UX tools are a \textit{hub} of collaborative activity for the entire team, not just among UXPs. Here, we use ``hub'' to refer to a common space where stakeholders from multiple domains can (and are invited to) collectively contribute, deliberate, and glean pertinent information, as opposed to space accessible to those from a single domain (e.g., UX). Similarly, we use ``hub tools'' to refer to tools that offer a hub workspace. For example, product managers can enter hub tools to receive progress updates and synthesize brainstormed ideas into higher-level product roadmaps, and developers can frequently reference designs in the tools as a guide during implementation and request design changes based on technical feasibility. 


That said, we also observed in Section \ref{s:collab-init} an imbalance in how collaborations are initiated. Despite heavy cross-functional collaborative activity in UX tools, most collaborations were initiated by UXPs. 
Additionally, we see in Section \ref{s:collab-stage} that the two roles with which UXPs collaborate the closest throughout the entire design process are UX designers and product managers. The lack of sustained collaboration with other roles is an indication that UXPs are typically not situated in a dominant role within a collaborative network (a ``hub'' role \cite{zhang2020data}) where they form extensive mutual collaborations with a wide range of non-UX roles \cite{calefato2016hub}. Instead, product managers appear to be the most promising candidates for hub roles, which is unsurprising given the cross-functional nature of the job \cite{chisa2014evolution}, but we cannot verify this hypothesis with our UXP-specific data. The combination of these two insights suggests an interesting distinction between the hub-like nature of \textit{UX tools} and \textit{UXPs}---the tools may act as a collaboration hub in the absence of the practitioners doing the same. 
In prior work on remote software teams,  researchers found that workers have become familiar with collaborative technologies \cite{bjorn2014distance} and thus have few issues with collaborative technology readiness, even as they continue to struggle with establishing common ground and bi-directional collaboration readiness \cite{olson2000distance}.
Thus, challenges with collaborative practices, such as miscommunication between UXPs and developers, may be harder to eradicate even as collaborative UX tools proliferate among non-UXPs.


UX tools are not the only example of hub tools used among software teams. Task management and communication platforms such as Jira, GitHub, and Slack also display similar characteristics \cite{chiu2002}, and we see increasing interest in \textit{inter-hub compatibility} where hubs offer and/or accept specialized integrations with other hubs \cite{jira, asana, figma-integrations}. \add{This is especially important when teams in different domains are already deeply accustomed to domain-specific representations and tools \cite{berg-tweet}.} In the context of multiple interconnected hubs, the aforementioned distinction suggests that hub tools should offer specialized features for hub roles to move horizontally between hubs and coordinate work efficiently. Strategies for using hub tools may also shift with inter-hub integrations---Hu et al. showed that tools and practices that benefit the inter-team collaboration may in fact harm intra-team collaboration, and vice versa \cite{hu2022paradox}. Whether and this phenomenon extends to UX or UX-adjacent hubs is a question ripe for investigation in future work.  


\subsection{Design Systems Require Governance and Flexibility}
The emergence of design systems has supported the scaling up of UX practices across diverse software products and services \cite{ms-ds, atlassian-ds, google-ds, uber-ds, ibm-ds}. \add{With this in mind, we asked RQ4: ``what are UXPs' current expectations and practices around reusing design components and managing design systems in collaborative UX tools?''} Design systems act on the promise of reducing effort, scaffolding learning, and increasing cross-functional collaboration \cite{churchill2019scaling}, but we found that many current design systems fail to effectively nurture collaborative behaviours and consequently undermine this promise. Specifically, there is little guidance on who is responsible for maintaining and enforcing the design system, which erodes its potential to be trusted and adopted. The lack of flexibility in the design system to account for new technical capabilities and interactions also limits its ability to scale. 

\subsubsection{Governance Handbook} Besides the 3 core components of design systems defined by Moore et al. \cite{moore2020systems} (see Section \ref{s:rel-works-tools}), we argue that a fourth component---a \textit{governance handbook}---should be included to eliminate ambiguity over who should maintain the design system and how they can do so. This handbook may include a system for assigning UXPs to oversee parts of the design system they work with as moderators, procedures for submitting and modifying components, guidelines for asking moderators questions about general usage, as well as guidelines for moderators to resolve design and usage inconsistencies that arise. Just like how a system of governance is essential for the continuity of a scalable, open source codebase \cite{latteman2005open}, a governance handbook can likewise help manage a scalable design system.

\subsubsection{Variation Patterns} In addition to a governance handbook, we recommend that \textit{interaction patterns} \cite{moore2020systems} specify not only reusable components and their interactions, but also \textit{variation patterns} that expose flexibility in their specifications. That is, not only does a component define its appearance and behaviour, but also how that appearance and behaviour may be adapted for new scenarios. For example, a variation pattern may include an extended font family to account for scenarios that demand sophisticated text formatting, and a guide on extending a button style to any UI featuring colour blocks, such as a progress bar. 
Variation patterns can help scale design systems into unforeseen territory by allowing practitioners to more concretely envision how existing patterns can be adapted and extended.

\subsection{Implications for UX Tools}
\subsubsection{One Tool, Multiple Modes}
One of our prominent findings is that Figma is strikingly popular with UXPs throughout all stages of the design process. We observed this across organization and UX team sizes. 
However, we also witnessed challenges arise from inviting collaborators into Figma's all-in-one canvas environment---developers 
found themselves wading through a trove of irrelevant screens in a design file they were not familiar with, leading to confusion. While using an open canvas that showcases evolution of work enhances process transparency, that same transparency can be overwhelming and obstructive when unnecessary. 

\begin{figure*}
    \centering
    \includegraphics[width=.9\textwidth]{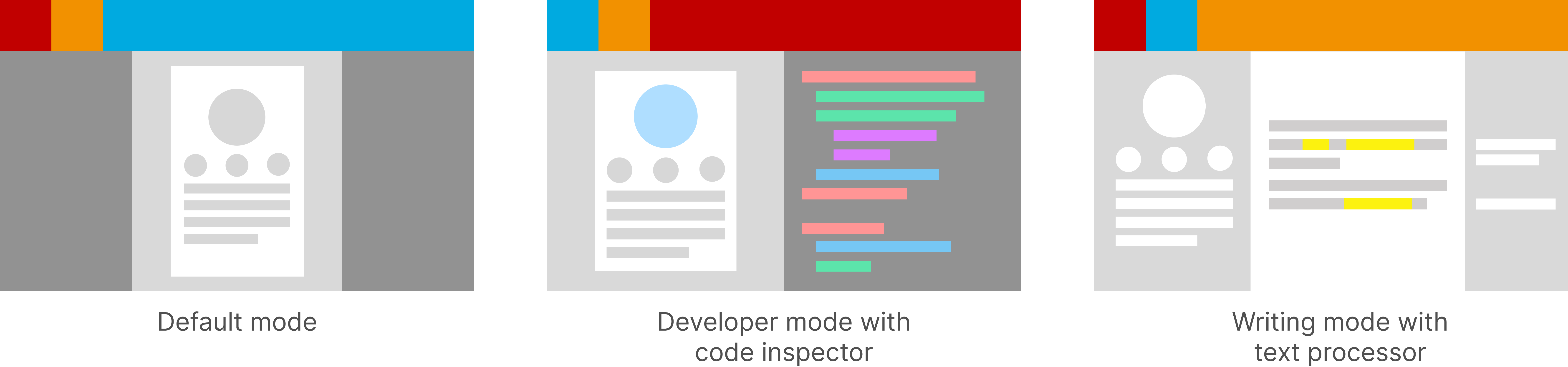}
    \caption{\add{Design tools can offer multiple modes for collaborating with different roles. For example, a designer may switch to ``developer mode'' and allow developers to experiment with code snippets alongside the UI, or ``writing mode'' to allow writers to focus on writing and editing UI text in a separate interface.}}
    \label{f:f-531}
\end{figure*}

Many design tools already support a ``presentation mode''\footnote{An example of such a mode is Figma's prototyping mode \cite{figma-prototype-mode}.} in which UXPs can present a polished, interactive prototype. This mode takes place in an environment separate from the rest of the design file and typically consists of an interface that mimics an end-user's interaction with the prototype. Our findings suggest that in addition to a presentation mode, \remove{other modes can also benefit collaboration} \add{design tools can introduce other modes to benefit collaboration}. Ideally, these modes will offer two capabilities: customization by users who initiate their use (e.g., selecting specific artifacts to show to specific collaborators) and supporting the attachment of mode-specific tools. For example, users can launch a ``developer mode'' in which only certain screens they select are shown to developers, accompanied by a web inspector to help extract HTML/CSS snippets. When working with UX Writers, a mode featuring an interface for commenting on and critiquing text may be more helpful than a standard design environment. \add{Wireframes of example modes are shown in Fig. \ref{f:f-531}.} By shifting to a multimodal approach in which only certain portions of the overall work environment are exposed at once, all-in-one tools can still offer the convenience and benefits of supporting work across design stages while improving organization during multi-role collaboration.

\subsubsection{Supporting Linkage of Design and Code Components}
\add{In RQ3, we asked, ``What practices and challenges arise when UXPs hand off their work to software developers and ML practitioners?''} We see from Section \ref{s:handoff} that despite efforts to establish clear channels of communication between UXPs and software developers, design-development divergences are still common. Previous work \cite{maudet2017breakdowns, leiva2019enact, walny2020handoff} have called upon co-creation practices to allow designers and developers to collaborate throughout the entire design process to minimize risk of divergence, but those practices do not seem to be actively used among UXPs in our population. Indeed, while 88 participants claimed they \textit{shared} their work with developers for handoff, at most 46 \textit{actively collaborated} with developers at any given stage.

\begin{figure*}
    \centering
    \includegraphics[width=1\textwidth]{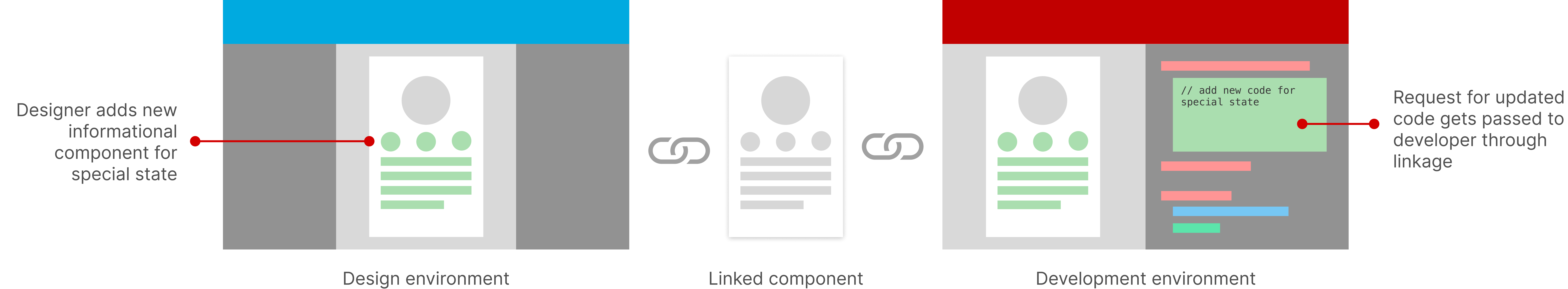}
    \caption{\add{A designer can manipulate a linked component via graphical editing in their design environment, while a developer can edit via code in their development environment. The link may also facilitate decision-making deliberations between the two parties.}}
    \label{f:f-532}
\end{figure*}

How can tools help UXPs think of handoffs as a co-creation effort rather than simply a transfer of work? One approach is to keep a design and its associated implementation mutually updated. That is, we propose \textit{linked components} where designers and developers can 1) design and prototype components through graphical manipulation, 2) actualize graphical components' interactive behaviours using code, and 3) collaboratively link the two and test for performance and edge cases \add{(see Fig. \ref{f:f-532})}. These links should be robust but flexible---changes to the component, whether it's the design or the code, can be indicated in the links and reviewed by others if necessary. \add{An example of this link in action is as follows. A designer mocks up a UI component that displays the results of a rider-driver match for a ride-hailing app. The developer adds frontend code to the component that implements its specified behavior. After some user testing, the designer realizes the empty state (in which there are no available drivers for the rider) is not handled well, and modifies the design accordingly. The change is signalled to the developer, who then updates the code for the empty state.} 
We see value in component-based links due to the rise of component-based design systems in UX \cite{churchill2019scaling}, a shift towards frontend programming frameworks emphasizing reusable components \cite{vue, react, svelte}, and a long-standing emphasis on unit testing in software engineering \cite{sen2005unit}. With our proposed approach of flexibly linking design and code components, we \textit{anticipate} design-development divergence and \textit{prepare practitioners for deliberation} rather than embarking on the much more arduous journey of eliminating all possibilities of divergence at all times.

\subsubsection{Writing assistance for on-design annotations}
\begin{figure*}
    \centering
    \includegraphics[width=1\textwidth]{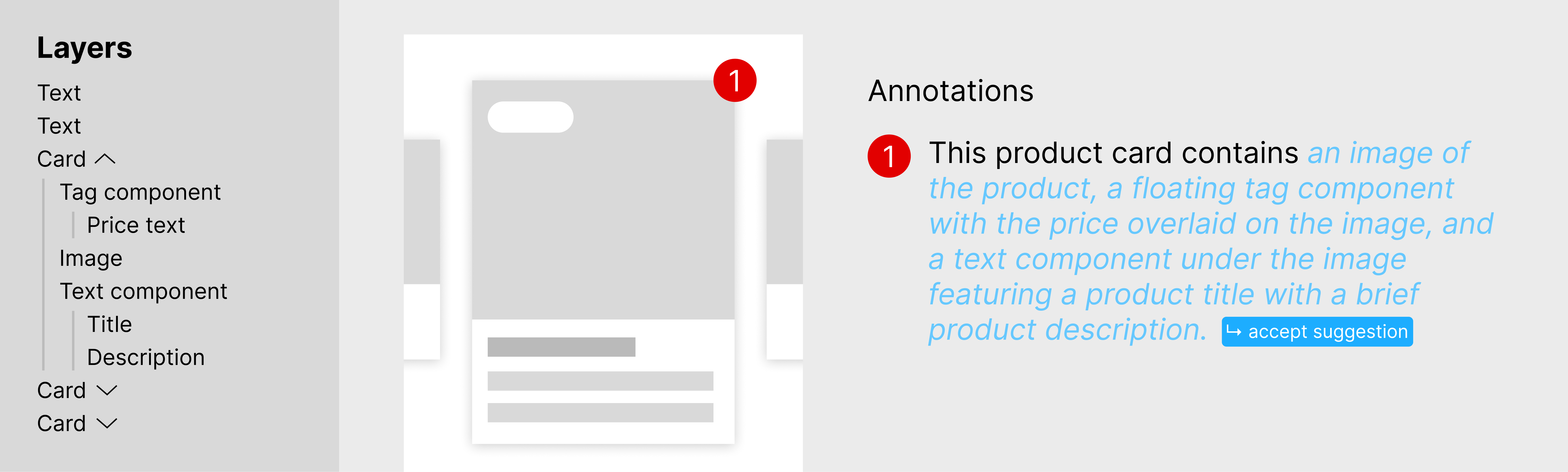}
    \caption{\add{A user starts typing an annotation. The AI-assisted writing tool then uses context from the user's snippet and combines it with information in the layer hierarchy tree in the left panel to provide writing suggestions for the rest of the annotation. The user can then press a key to accept the suggestion, or ignore it.}}
    \label{f:f-533}
\end{figure*}

Despite robust commenting capabilities in collaborative tools, we found that UXPs still heavily rely on writing text directly on or beside their designs to document design decisions, citing superior visibility over comments as a reason for doing so. Writing these annotations is often an open-ended---and in the case of larger projects, daunting and time-consuming---task, especially if there are few guidelines on how to discuss design decisions, how much detail to write, and whether a common structure or template should be used for different annotations. 
Conveniently, there has been a surge in interest in AI-assisted writing with large language models for tasks such as creative writing \cite{chung2022talebrush, Calderwood2020HowNU, swanson2021story}, scientific writing \cite{gero2022sparks, merono2020can}, and probing the model's capabilities as a writing collaborator \cite{lee2022coauthor}. These writing tasks share the same open-ended nature (and in the case of scientific writing, the logical nature as well) of writing annotations. In addition to models working purely with text, multimodal transformers such as CLIP \cite{radford2021clip} can learn representations that link textual data with visual data 
\cite{openai-clip}. Multimodal transformers with AI-assisted writing present a powerful combination for authoring on-design annotations. References to visual components can be (optionally) translated to text for writing inspiration, and writing of the annotation itself can be further enhanced by suggestions from language models. This approach can draw upon work from the accessibility community on screen readers, which already translate visual web components to descriptive text \cite{morris2018screen}. \add{The layer hierarchy tree in design tools can also be used by the model to learn relationships between components. The learned UI structure can then be supplemented user-provided context to automatically generate semantically rich, context-aware annotations. As an example (Fig. \ref{f:f-533}), when a user starts typing an annotation, an intelligent agent takes context from that snippet and, leveraging information about UI relationships given by the layer hierarchy tree in the left panel, suggests an autocompletion for the annotation.}

Note that writing assistance for annotations may not necessarily be generalizable to comments. Annotations are \textit{explanatory} while comments are more \textit{prescriptive}---commentors typically communicate a message about an artifact or concept rather than explain the artifact or concept itself. We defer the further disambiguation of annotations and comments to future work. 

\section{Limitations and Future Work}
Our study consisted of only UXPs, which meant we were only able to capture one side of the narrative when it comes to cross-functional collaboration. For example, for developer handoff, perspectives of developers who work with UXPs in implementing designs would aptly supplement our study. Similarly, perspectives from product managers can help us verify our hypothesis that product management is a hub role within software teams. Surveying non-UXPs who work with UXPs may also allow reciprocity in collaboration \cite{zhang2020data} (or lack thereof) to be surfaced. As such, surveying non-UXPs about their practices and tools when collaborating with UXPs is a promising avenue for future work. 

We also observed a sample bias towards large organizations of 10,000+ employees. Large organizations may have UX practices and tools that differ from smaller organizations \cite{gray2015}, and although we observed hints of those differences in Section \ref{s:collab-stage}, they may be dampened by the relatively low number of participants not employed by large organizations. 
To tease apart differences in practices across organization sizes, future work can recruit UXPs who worked at organizations of various sizes for an interview study about their collaboration experiences. 
Additionally, to protect confidentiality around specialized procedures that may be in use at particular organizations, we did not collect names of employers from our participants and therefore could not identify employer-related biases in our responses. 

\add{Although our survey was generally effective in surfacing both high-level practices across the design process as well as more granular practices within each stage, it has its limitations. The survey abstracted the UX process to a linear workflow to align with the ubiquitous double diamond model, but in practice, iterative loops can exist between any and all of the stages. Future versions of our survey can focus on those loops to investigate collaboration challenges when iterating on UX work. Moreover, the balance of breadth and depth in our survey meant that some of our inquiries generated rich responses that we would have liked to further investigate, while others were not as relevant. For example, we felt design systems management was an area that could use more collaborative support given participants' experiences and asking more questions oriented around this topic could have been more valuable, but our section on design environment customization did not reveal any pertinent findings and we did not end up reporting on it. We can use the lessons learned in this survey to inform the design of future surveys, as well as follow-up interview and co-design studies, to yield more salient insights.}

Finally, this study focused on software organizations based in the U.S. to ensure consistent organizational understanding of UX \cite{lallemand2015}. Future work can compare and contrast UX collaboration practices across countries and continents to strengthen our understanding of UX as a global field.

\section{Conclusion}

We deployed a survey to professional UXPs in software organizations to better characterize their collaborative practices and use of tools. We found that UXPs actively collaborated with team members in shared, multiplayer environments such as Figma across all stages of the design process, usually initiating the collaborations themselves. These environments offer new ways for UXPs to communicate and receive feedback on their work; however, they also present novel possibilities for collaborative friction. We posit that tool multimodality, component linkage between design and code, and assistance for authoring on-design text annotations can mitigate this friction and improve collaborative potential of UX tooling. Tool enhancement alone, however, is not a panacea. New collaborative artifacts and practices, such as a governance handbook for design systems, need to be considered alongside tools for a holistic approach to collaboration enrichment.

UX today is experiencing pivotal growth in terms of the number of practitioners, adoption by organizations, and accessibility to non-UXPs. In the midst of this growth, our work illuminates potential paths forward for advancing collaboration within and around UX to streamline the design of human-centered, interactive systems in practice.


\begin{acks}
We extend a warm thanks to all of our participants and reviewers. We also thank Cristen Torrey and Joy Kim for their conversations and early feedback on this project. 
\end{acks}

\bibliographystyle{ACM-Reference-Format}
\bibliography{refs}



\end{document}